\documentclass[english,aps,prl,twocolumn,amsmath,amssymb,showpacs,superscriptaddress,notitlepage,longbibliography]{revtex4-1}
\usepackage[T1]{fontenc}
\usepackage[latin9]{inputenc}
\usepackage{amsmath}
\usepackage{amssymb}
\usepackage{graphicx}
\usepackage{mathrsfs}
\makeatletter
\providecommand{\tabularnewline}{\\}
\usepackage{amssymb}
\usepackage{amsmath}
\usepackage{graphicx}
\usepackage[colorlinks=true,linkcolor=blue,anchorcolor=red,citecolor=blue,urlcolor=blue]{hyperref}
\usepackage[caption=false]{subfig}
\usepackage{lipsum}
\usepackage{balance}

\usepackage{ulem}
\makeatother
\usepackage{babel}
\begin{document}
\renewcommand{\figurename}{Fig.}
\title{Intrinsic finite-energy Cooper pairing in $j=3/2$ superconductors}
\author{Masoud Bahari}
\email{masoud.bahari@physik.uni-wuerzburg.de}

\affiliation{Institute for Theoretical Physics and Astrophysics$\text{,}$ University
of Würzburg, D-97074 Würzburg, Germany}
\affiliation{Würzburg-Dresden Cluster of Excellence ct.qmat, Germany}
\author{Song-Bo Zhang}
\affiliation{Institute for Theoretical Physics and Astrophysics$\text{,}$ University
of Würzburg, D-97074 Würzburg, Germany}
\affiliation{Würzburg-Dresden Cluster of Excellence ct.qmat, Germany}
\author{Björn Trauzettel}
\affiliation{Institute for Theoretical Physics and Astrophysics$\text{,}$ University
of Würzburg, D-97074 Würzburg, Germany}
\affiliation{Würzburg-Dresden Cluster of Excellence ct.qmat, Germany}
\date{\today}
\begin{abstract}
We show that Cooper pairing can occur intrinsically away from the
Fermi surface in $j=3/2$ superconductors with strong spin-orbit coupling
and equally curved bands in the normal state. In contrast to conventional
pairing between spin-$1/2$ electrons, we derive that pairing can
happen between inter-band electrons having different magnetic quantum numbers, for instance, $m_j=1/2$ and $m_j=3/2$. Such superconducting correlations manifest themselves by a pair of indirect gap-like structures
at finite excitation energies. An observable signature of this exotic
pairing is the emergence of a pair of symmetric superconducting coherence
peaks in the density of states at finite energies. Moreover, the angular-momentum-resolved density of states in the presence of a perturbative Zeeman field reflects the $m_j$ composition of the Cooper pairs. We argue that such finite-energy pairing is a generic feature of $j=3/2$ superconductors, both in presence and absence of inversion symmetry.
\end{abstract}
\maketitle
\textit{Introduction.}\textemdash Since the discovery of Bardeen\textendash Cooper\textendash Schrieffer theory for superconductivity
\citep{BCS}, extensive efforts of theoretical and experimental research
have been carried out to understand the pairing mechanism \citep{Rev2,ManfredSigrist}.
In most cases, superconductivity can be described by pairing of spin-$1/2$
electrons at the Fermi surface. However, it has been shown theoretically
that pairing of electrons with higher  total angular momentum is also
possible \citep{ArbitrarySpin,HighPartial,Brydon}. This has triggered
attempts to formulate a general theory of high angular momentum superconductivity
\citep{Theo2,Theo1,Theo3,Annica} and to identify typical physical
observables \citep{TriDirac,Igor,Bogo0,InflatednodeYPtBi,Mixing,J32TopoInsu,MultiPoleRes,Bitan,Bogo1,Bogo3,Bogo2,Bitan2}.
Prominent candidate materials for high angular momentum superconductivity are
half-Heusler compounds whose Fermi surface lies close to the $\Gamma_{8}$
band with total angular momentum quantum number $j=3/2$ \citep{BiSup5,Nature1,Nature3,YPtBi1,YPtBi2,BiSup3,BiSup4,BiSup13,BiSup1, BiSup2,Nature2,YPtBi3,BiSup7,LondonPent,BiSup6,BiSup14,BiSup8,BiSup9,BiSup12}.
These materials can be categorized into two distinct groups with inverted
\citep{BiSup5,Nature1,Nature3,YPtBi1,YPtBi2,BiSup3,BiSup4,BiSup13,BiSup1, BiSup2,Nature2,YPtBi3,BiSup7,LondonPent,BiSup14,BiSup8,BiSup9,BiSup12,ArxSeptEx}
and normal \citep{BiSup1,BiSup6,BiSup12,BiSup14} band structures, respectively.
In the inverted case, only a single pair of $\Gamma_{8}$ bands with
identical components of total angular momentum cross the Fermi energy \citep{DFT1,ArxSeptEx,MatProj,DFT2,Brydon}.
Despite the $j=3/2$ nature of the electrons, the pairing mechanism in
this case can be captured within the formalism for (pseudo)spin-$1/2$
electrons at low energies \citep{Brydon}. In contrast, in the group
with normal band structure, density functional theory calculations
predict that all $\Gamma_{8}$ bands bend downward near
the Fermi energy \citep{MatProj,Nature1,Nature3,DFT1,DFT2,BiSup12}.
This band structure applies, for instance, to \textit{R}PdBi with
$R\in$\{Y, Dy, Tb, Sm\} \citep{BiSup1,BiSup6,BiSup12,BiSup14,CmFermi}. We demonstrate below that such configuration
of energy bands in combination with superconductivity allow us to
observe Cooper pairing composed by electrons with non-identical magnetic quantum numbers $m_j$ at finite excitation energies (FEE).

Pairing of spin-1/2 electrons with different orbitals at FEE has been proposed for the material MgB$_2$ in absence of spin-orbit coupling \citep{Moreo}. Recently, it has been argued that Ising superconductors may realize
finite-energy pairing of spin-1/2 electrons by applying external in-plane
magnetic fields \citep{Mirage}.

Hence, the novel question we address in this Letter is whether it is possible to observe intrinsic finite-energy Cooper pairing composed by electrons with \textit{different} magnetic quantum numbers in the absence of any fields. We show below that the interplay of strong spin-orbit coupling and superconductivity allows for such pairing accompanied by a pair of indirect \textit{gap-like structures} (GLSs) away from the Fermi energy. The electrons responsible
for the finite-energy pairing originate from energy bands with different
band indices. Our results suggest that
such behavior is a generic feature of multiband superconductors when the $j=3/2$ electrons of the $\Gamma_{8}$ band contribute to pairing.
In experiments, the GLSs manifest themselves by the appearance of
a pair of symmetric superconducting coherence peaks at FEE of the density of states (DOS). To elucidate that such novel Cooper pairing is a generic phenomenon of multi-band superconductors
preserving (breaking) inversion symmetry, we systematically analyze
the role of $j=3/2$ pairing valid for cubic point group symmetry
$O_{h}$ ($T_{d}$) based on the Luttinger-Kohn model.

\textit{Model.}\textemdash Low-energy $j=3/2$ electrons within the
$\Gamma_{8}$ bands can be described by the \textbf{k$\cdot$p} Luttinger-Kohn
model \citep{Luttinger,Dresselhaus}, $H_{0}=\sum_{{\bf k}}\hat{c}_{{\bf k}}^{\dagger}\mathcal{\hat{H}}_{0}({\bf k})\hat{c}_{{\bf k}}$,
where
\begin{align}
\mathcal{\hat{H}}_{0}({\bf k})= & \alpha k^{2}\hat{I}_{4}+\beta\sum_{i}k_{i}^{2}\hat{J}_{i}^{2}+\gamma\sum_{i\neq j}k_{i}k_{j}\hat{J}_{i}\hat{J}_{j}-\mu\hat{I}_{4},\label{NormalH}
\end{align}
and the basis is $\hat{c}_{{\bf k}}=(c_{{\bf k},3/2},c_{{\bf k},1/2},c_{{\bf k},-1/2},c_{{\bf k},-3/2})^{T}$.
We denote ${\bf k}=(k_{x},k_{y},k_{z})$ as the 3D momentum, $k=|{\bf k}|$,
$\hat{J}_{i}$ with $i\in\{x,y,z\}$ as the $4\times4$ total angular
momentum matrices in $j=3/2$ representation, and $\hat{I}_{4}$ as
the $4\times4$ identity matrix. The material-dependent parameters
$\alpha$ and $\beta$ $(\gamma)$ control kinetic energy and symmetric
spin-orbit coupling, respectively; $\mu$ is the Fermi energy. The
doubly-degenerate eigenenergies of $\mathcal{\hat{H}}_{0}({\bf k})$,
protected by the combination of inversion and time-reversal symmetries,
are given by
\begin{equation}
E_{{\bf k}}^{\pm}\!=\!\!\left(\!\!\alpha\!+\!\frac{5}{4}\beta\!\!\right)\!k^{2}\!\pm\!\beta\!\sqrt{\sum_{i}\!\left[k_{i}^{4}\!+\!\!\!\left(\frac{3\gamma^{2}}{\beta^{2}}-\!1\!\right)\!k_{i}^{2}k_{i+1}^{2}\!\right]}-\mu,\label{Dispersion}
\end{equation}
where $i+1=y$ if $i=x$ (notation used throughout the paper). To
investigate the properties of the excitation spectrum of Eq.\ (\ref{NormalH})
in the presence of high angular momentum Cooper pairing in $O_{h}$ symmetry, we introduce the full superconducting Hamiltonian given
by $H=\sum_{{\bf k}}\hat{\psi}_{{\bf k}}^{\dagger}\hat{H}_{\text{BdG}}({\bf k})\hat{\psi}_{{\bf k}}$,
where $\hat{\psi}_{{\bf k}}=(\hat{c}_{\mathbf{k}},\hat{c}_{-\mathbf{k}}^{\dagger T})^{T}$
is the Nambu spinor. The Bogoliubov-de Gennes (BdG) Hamiltonian takes
the form
\begin{equation}
\hat{H}_{\text{BdG}}({\bf k})=\left(\begin{array}{cc}
\hat{\mathcal{H}}_{0}({\bf k}) & \ \ \hat{\mathcal{H}}_{\eta}^{J,S,L}({\bf k})\\{}
[\hat{\mathcal{H}}_{\eta}^{J,S,L}({\bf k})]^{\dagger} & -\hat{\mathcal{H}}_{0}^{T}(-{\bf k})
\end{array}\right),
\end{equation}
where $\hat{\mathcal{H}}_{\eta}^{J,S,L}({\bf k})$ is the pairing
Hamiltonian in channel $(\eta,J,S,L)$ with $\eta$ being the relative
basis label of the cubic irreducible representation (IR) \citep{GroTinkham,GroDress}.
The channel of instability is named by Cooper pair quantum numbers
with total angular momentum $J$ combining intrinsic spin $S$ and
orbital $L$ angular momenta \citep{Theo1,Theo3,Supp}.
\begin{figure}
\begin{centering}
\includegraphics[scale=0.36]{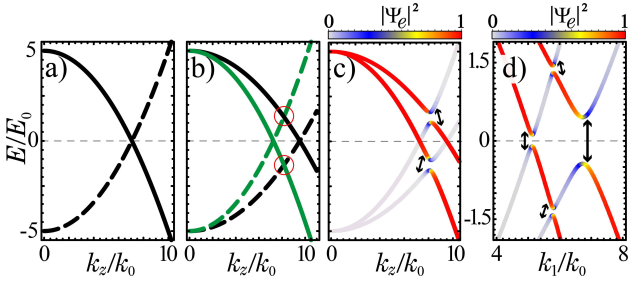}
\par\end{centering}
\caption{\label{Fig1} BdG spectra along the $[0,0,1]$
direction in absence of pairing for (a) $\beta=0$ and (b) $\beta=0.2|\alpha|$,
respectively. The spectra are independent
of $\gamma$. BdG spectrum in the (c) $[0,0,1]$ and (d) $[1,1,0]$ (i.e., $k_{x}=k_{y}=k_{1}$) directions for $\gamma=\beta$ in presence of septet pairing with amplitude $\Delta/E_{0}a=4.15$. The color denotes
the probability of electronic states $|\Psi_{e}|^{2}$ in both (c)
and (d) panels. Other parameters are $\mu/E_{0}=-5$, $k_{0}=10^{-2}a^{-1}$,
$E_{0}=10^{-3}|\alpha|a^{-2}$ and $\alpha=-20$. $a$ is the lattice constant in a tight-binding version of the continuum
model.}
\end{figure}

To shed light on finite-energy pairing, the BdG excitation
spectrum along the $[0,0,1]$ direction in absence of spin-orbit coupling
and pairing is plotted in Fig.\ \ref{Fig1}(a). The fourfold degenerate
electron bands (solid line) cross their hole counterparts (dashed
line) at $k_{F}=\sqrt{\mu/\alpha}$. A finite $\beta$ accounting
for spin-orbit coupling splits the energy bands having different magnetic quantum numbers. Increasing $\beta$, this moves the crossings at the Fermi
surface $E=0$ and at FEE (red circles), as shown in Fig.\ \ref{Fig1}(b).
The low-energy $m_j$ split Fermi momenta consist of $m_j=1/2$ states
(green) and $m_j=3/2$ states (black) located at $k_{F}^{-}=2\sqrt{\mu/(4\alpha+\beta)}$
and $k_{F}^{+}=2\sqrt{\mu/(4\alpha+9\beta)}$, respectively. Moreover,
the finite-energy crossing appears at $\tilde{k}=2\sqrt{\mu/(4\alpha+5\beta)}$
incorporating $m_j=3/2$ electron (hole) and $m_j=1/2$ hole (electron)
states at positive (negative) excitation energies. In the superconducting
state, the pairing mechanism occurs not only at $E=0$ but also at
FEE {[}Figs.\ \ref{Fig1}(c) and \ref{Fig1}(d){]}. Notably, the
finite-energy pairing can be present when the low-energy intra-band
states exhibit nodal {[}Fig.\ \ref{Fig1}(c){]} or gapped excitation
spectra {[}Fig. \ref{Fig1}(d){]}.

\textit{Finite-energy effective theory.}\textemdash To better understand
the finite-energy pairing, we develop an effective theory close to
the FEE. We start by obtaining the band basis representation of the
BdG Hamiltonian through the basis transformation $\hat{c}_{{\bf k}}=\hat{V}_{{\bf k}}^{+}\hat{f}_{{\bf k}}^{+}+\hat{V}_{{\bf k}}^{-}\hat{f}_{{\bf k}}^{-}$,
where $\hat{V}_{{\bf k}}^{\pm}$ is a $4\times2$ matrix containing
the eigenvectors corresponding to $E_{{\bf k}}^{\pm}$. Note that
$\hat{f}_{{\bf k}}^{\pm}=(f_{{\bf k},\uparrow}^{\pm},f_{{\bf k},\downarrow}^{\pm})^{T}$
and $f_{{\bf k},s}^{\pm}$ ($f_{{\bf k},s}^{\pm\dagger}$) annihilates
(creates) a state with pseudospin degrees of freedom $s\in\{\uparrow,\downarrow\}$
in the band basis labeled by $\pm$ in Eq.\ (\ref{Dispersion}).
To capture the inter-band superconducting Hamiltonian, we choose our
basis set as $\hat{\varphi}_{{\bf k}}=(\hat{\varphi}_{{\bf k}}^{+-},\hat{\varphi}_{{\bf k}}^{-+})^{T}$ with $\hat{\varphi}_{{\bf k}}^{+-}=(\hat{f}_{{\bf k}}^{+},(\hat{f}_{-{\bf k}}^{-\dagger})^{T})^{T}$ denoting the electron-hole subspace basis with band index $(+,-)$
and $\hat{\varphi}_{{\bf k}}^{-+}=(\hat{f}_{{\bf k}}^{-},(\hat{f}_{-{\bf k}}^{+\dagger}){}^{T})^{T}$. Thus, we rewrite
the superconducting Hamiltonian in the band basis as $H=\sum_{{\bf k}}\hat{\varphi}_{{\bf k}}^{\dagger}\hat{h}({\bf k})\hat{\varphi}_{{\bf k}}$
with
\begin{equation}
\hat{h}({\bf k})=\left(\begin{array}{cccc}
E_{{\bf k}}^{+} & \hat{\Delta}_{{\bf k}}^{+-} & 0 & \hat{\Delta}_{{\bf k}}^{++}\\
(\hat{\Delta}_{{\bf k}}^{+-})^{\dagger} & -E_{{\bf k}}^{-} & (\hat{\Delta}_{{\bf k}}^{--})^{\dagger} & 0\\
0 & \hat{\Delta}_{{\bf k}}^{--} & E_{{\bf k}}^{-} & \hat{\Delta}_{{\bf k}}^{-+}\\
(\hat{\Delta}_{{\bf k}}^{++})^{\dagger} & 0 & (\hat{\Delta}_{{\bf k}}^{-+})^{\dagger} & -E_{{\bf k}}^{+}
\end{array}\right),\label{interand}
\end{equation}
where $\hat{\Delta}_{{\bf k}}^{+-}$ is the projection of the pairing
instability onto the inter-band basis given by $\hat{\Delta}_{{\bf k}}^{+-}=\hat{V}_{{\bf k}}^{+\dagger}\hat{\mathcal{H}}_{\eta}^{J,S,L}({\bf k})(\hat{V}_{-{\bf k}}^{-\dagger})^{T}$.
Treating the off-diagonal blocks, corresponding to the intra-band
pairing denoted by $\hat{\Delta}_{{\bf k}}^{\nu\nu}$ with $\nu\in\{+,-\}$,
as a perturbation to the inter-band diagonal block and employing the
folding down approach \citep{Fold}, we arrive at the effective Hamiltonian valid in the vicinity of the GLSs
\begin{equation}
H_{\text{eff}}^{+-}({\bf k})\!=\!\left(\!\begin{array}{cc}
E_{{\bf k}}^{+}\!+\!\hat{\varepsilon}_{{\bf k}}^{++} & \hat{\Delta}_{\text{eff}}^{+-}({\bf k})\\
\big(\hat{\Delta}_{\text{eff}}^{+-}({\bf k})\big)^{\dagger} & -E_{{\bf k}}^{-}\!+\!\hat{\varepsilon}_{{\bf k}}^{--}
\end{array}\!\!\right).\label{Heff+-}
\end{equation}
The second term on the diagonal in Eq.\ (\ref{Heff+-}) is a pseudospin
energy shift induced by the pairing of intra-band quasi-particles,
given by $\hat{\varepsilon}_{\bf k}^{\nu\nu}=\hat{\Delta}_{\bf k}^{\nu\nu}(\hat{\Delta}_{\bf k}^{\nu\nu})^{\dagger}/(\omega+\nu E_{\bf k}^{\nu})$. Notably, Eq.\ (\ref{Heff+-}) is different
from a typical BdG Hamiltonian. The effective particle-hole symmetry
is broken due to the presence of non-identical diagonal entries arising
from the nature of two different energy bands. The inter-band pairing of the effective Hamiltonian takes the form
\begin{equation}
\hat{\Delta}_{\text{eff}}^{+-}({\bf k})=\hat{\Delta}_{{\bf k}}^{+-}+\varepsilon_{{\bf k}}^{-1}\hat{\Delta}_{{\bf k}}^{++}(\hat{\Delta}_{{\bf k}}^{-+})^{\dagger}\hat{\Delta}_{{\bf k}}^{--},
\end{equation}
where $\varepsilon_{{\bf k}}=(\omega+E_{{\bf k}}^{+})(\omega-E_{{\bf k}}^{-})$
\citep{cm7}. In the weak-pairing limit, the second term is small
close to the GLSs and can be neglected. The spectrum
for the FEE reads
\begin{equation}
\mathcal{E}_{\pm}({\bf k})=\varepsilon_{{\bf k},1}+\varepsilon_{{\bf k},2}\pm\sqrt{(\varepsilon_{{\bf k},1}-\varepsilon_{{\bf k},2})^{2}+\mathring{\delta}({\bf k})},
\end{equation}
where
\begin{equation}
\mathring{\delta}({\bf k})=\frac{1}{2}\text{Tr}(\hat{\Delta}_{\text{eff}}^{+-}({\bf k})[\hat{\Delta}_{\text{eff}}^{+-}({\bf k})]^{\dagger}),\label{interbandStrength}
\end{equation}
is the magnitude of the GLS indicating superconducting hybridization
between inter-band states \citep{cm5}, i.e., pairing of $m_j=3/2$ with $m_j=1/2$ states; $\text{Tr}$ stands for the trace of the matrix;
$\varepsilon_{{\bf k},1}=(1/2)E_{{\bf k}}^{+}+(1/4)\text{Tr}(\hat{\varepsilon}_{{\bf k}}^{++})$
and $\varepsilon_{{\bf k},2}=-(1/2)E_{{\bf k}}^{-}+(1/4)\text{Tr}(\hat{\varepsilon}_{{\bf k}}^{--})$.
The width of the GLSs around the finite-energy crossing momentum \citep{PreqGLS}
is $|\mathcal{E}_{+}(\tilde{\boldsymbol{k}})-\mathcal{E}_{-}(\tilde{\boldsymbol{k}})|=2[\mathring{\delta}(\tilde{\boldsymbol{k}})]^{1/2}$.
Note that the matrix form of $\hat{\Delta}_{\text{eff}}^{+-}({\bf k})$
depends on the choice of basis while $\mathring{\delta}({\bf k})$
is a basis-independent observable.

\begin{figure}
\begin{centering}
\includegraphics[scale=0.34]{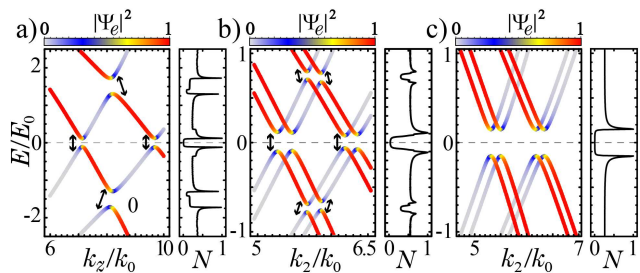}
\par\end{centering}
\caption{\label{Fig2} BdG spectra in (a) $[0,0,1]$ and (b and c) $[1,0,1]$
(with $k_{x}=k_{z}=k_{2}$) directions for (a) $(\Delta/E_{0}a,\delta/E_{0}a)=(3.5,10)$,
(b) $(1.5,2.5)$, and (c) $(0.15,2.5)$, respectively. (a) and (b)
correspond to septet pairing, and (c) corresponds
to $A_{1g}$ pairing. The corresponding density
of states $N$ normalized with respect to its maximum value are presented
in the right panel of each spectrum. $\gamma=-0.05|\alpha|$ and other
parameters are the same as those in Fig.\ \ref{Fig1}.}
\end{figure}

\textit{Symmetry properties.}\textemdash Interestingly, the symmetry
properties of the finite-energy pairing are different from their low-energy
counterpart. For instance, we may witness even(odd)-parity pseudo-spin
triplet (singlet) pairing at FEE. This is a direct consequence of
the Pauli exclusion principle taking into account the exchange of
band-indices in addition to the exchange of magnetic quantum numbers, i.e.,
\begin{align}
\hat{\Delta}_{\text{eff}}^{+-}(-{\bf k}) & =-[\hat{\Delta}_{\text{eff}}^{-+}({\bf k})]^{T}.\label{InterbandSymm}
\end{align}
In this sense, we can span $\hat{\Delta}_{\text{eff}}^{+-}({\bf k})$
in the inter-band basis as $\hat{\Delta}_{\text{eff}}^{+-}({\bf k})=\boldsymbol{\mathfrak{g}}^{+-}({\bf k}).\boldsymbol{\tau}$,
where the four-component vector $\boldsymbol{\mathfrak{g}}^{+-}=(\mathfrak{g}_{0}^{+-},\mathfrak{g}_{x}^{+-},\mathfrak{g}_{y}^{+-},\mathfrak{g}_{z}^{+-})$
is a complex momentum dependent function, $\boldsymbol{\tau}=(\tau_{0},\tau_{x},\tau_{y},\tau_{z})$
with $\tau_{x,y,z}$ being the Pauli matrices and $\tau_{0}$ the
$2\times2$ identity matrix in the inter-band basis. Thus, we obtain
the symmetry relations
\begin{align}
\mathfrak{g}_{0,x,z}^{+-}(-{\bf k}) & =-\mathfrak{g}_{0,x,z}^{-+}({\bf k}),\ \ \ \mathfrak{g}_{y}^{+-}(-{\bf k})=\mathfrak{g}_{y}^{-+}({\bf k}).\label{gvecotr}
\end{align}
This enables us to directly derive components of the $\hat{\Delta}_{\text{eff}}^{-+}({\bf k})$. The $y$-component
is even in momentum while the other components are odd
\citep{cm10}.

\textit{Pairing channels of $O_{h}$ symmetry.}\textemdash We apply
our theory to all time-reversal symmetric stationary pairing states
of cubic point group symmetry up to the $p$-wave channel \citep{Theo3,cm9}
with the aim to identify inter-band pairing. To obtain analytic relations
for $\mathring{\delta}({\bf k})$, we set $\gamma=\beta$ \citep{cm4}. Note that the pairing states generate cubic anisotropy. The results
are summarized in Table\ \ref{Table1}. Remarkably, inter-band pairing
is present for a variety of pairing channels.

\begin{table}[tb]
\begin{tabular}{ccccc}
\hline
$O_{h}$ ($T_{d}$) & $\eta$ & $(J,S,L)$ & without ASOC & with ASOC\tabularnewline
\hline
$A_{1g}$ ($A_{1}$) & $I$ & $(0,0,0)$ & $\times$ & $\times$\tabularnewline
$A_{1u}$ ($A_{2}$) & $f(\boldsymbol{r})$ & $(0,1,1)$ & $\times$ & $[1,1,1]^{*}$\tabularnewline
$T_{1u}$ ($T_{2}$) & $z$ & $(1,1,1)$ & $[0,0,1]$ & $[0,0,1]$\tabularnewline
$E_{g}$ ($E)$ & \ \ $3z^{2}-r^{2}$\ \  & $(2,2,0)$ & $[0,0,1]$ & $\checkmark$\tabularnewline
 & $x^{2}-y^{2}$ & $(2,2,0)$ & $\checkmark$ & $\checkmark$\tabularnewline
$E_{u}$ ($E$) & $3z^{2}-r^{2}$ & $(2,1,1)$ & \ $[0,0,1]^{*},k_{z}=0$ & $\checkmark$\tabularnewline
 & $3z^{2}-r^{2}$ & $(2,3,1)$ & $[0,0,1]$ & $\checkmark$\tabularnewline
 & $x^{2}-y^{2}$ & $(2,1,1)$ & $[0,0,1]^{*}$ & $[0,0,1]$\tabularnewline
 & $x^{2}-y^{2}$ & $(2,3,1)$ & $\checkmark$ & $\checkmark$\tabularnewline
$A_{2u}$ ($A_{1}$) & $xyz$ & $(3,3,1)$ & $[1,1,1]^{*}$ & $[1,1,1]^{*}$\tabularnewline
\hline
\end{tabular}

\caption{\label{Table1} Absence/presence of finite-energy Cooper pairing.
The first column shows the IR of $O_{h}$ ($T_{d})$ point groups
with $\eta$ denoting the basis label of the IR. The third column corresponds
to the pairing multiplets of the relative IR. The last two columns
indicate presence $\checkmark$ (absence $\times$) of finite-energy
Cooper pairing in the entire momentum space in absence and presence
of ASOC. The superscript ({*}) means that $\mathring{\delta}({\bf k})$
vanishes in all equivalent directions \citep{EqiDirec}.}
\end{table}

First, we observe that the even- and odd-parity singlet pairing
states \citep{ChanNaming}, corresponding to the instability channels $A_{1g}$ and $A_{1u}$,
respectively, have vanishing inter-band pairing, i.e., $\mathring{\delta}({\bf k})=0$
\citep{cm6}. Contrarily, the cubic triplet state $T_{1u}$ \citep{InertStates,Theo3}
shows finite inter-band pairing $\mathring{\delta}({\bf k})\!=\!\Delta^{2}(k_{x}^{2}+k_{y}^{2})$
with $\Delta$ being the pairing strength. This indicates that the
GLSs are present within the whole momentum space except for the $[0,0,1]$
direction where inter-band pairing vanishes. Next, we focus on pairing with quintet total angular momentum, i.e., $J=2$. In this
case, the pairing state is split by the cubic field into $E_{g,u}+T_{2g,u}$
where $E_{g,u}$ ($T_{2g,u}$) is a two(three)-dimensional IR. Note that the pairing state $E_{g,u}$ is a stationary state of the free energy whereas $T_{2g,u}$ is not \citep{Theo3}. Hence, we focus on $E_{g,u}$ pairing in the following. The components of $E_{g,u}$ are denoted by $\eta=(3z^{2}-r^{2},x^{2}-y^{2})$. In the $j=3/2$ representation, we find two (four) symmetry allowed
pairing channels for even-parity (odd-parity) quintet pairing.
For even-parity states, the quantum number is $(2,2,0)$, where
the pairing Hamiltonian is momentum independent due to the $s$-wave
nature of the channel. In this case, the GLSs of the $3z^{2}-r^{2}$
state are given by $\mathring{\delta}({\bf k})\!=\!3\Delta_{{\bf k}}^{2}(k_{x}^{2}\!+\!k_{y}^{2})(k^{2}\!+\!3k_{z}^{2})$
with $\Delta_{{\bf k}}=\Delta/2k$, showing non-vanishing GLSs except for the two-fold rotation axis $[0,0,1]$. Importantly, the $x^{2}-y^{2}$ state exhibits
full GLSs within the entire momentum space.

The odd-parity quintet channel has four momentum dependent stationary
pairing states due to $L=1$. The first two states correspond to the
$3z^{2}-r^{2}$ basis having Cooper pair quantum numbers $(2,1,1)$
and $(2,3,1)$. These states differ only in the intrinsic spin quantum
number where $S=1$ and $S=3$ denote spin dipole and octupole moments,
respectively. The GLS for the former state takes the form
$\mathring{\delta}({\bf k})\!=\!27\Delta_{k}^{2}\ (k_{x}^{2}+k_{y}^{2})k_{z}^{2}$. It vanishes in the $[0,0,1]^*$ direction \citep{EqiDirec} as well as the $k_{z}=0$ plane. For
the $S=3$ channel, the GLS becomes
\begin{equation}
\mathring{\delta}({\bf k})\!=\!\Delta_{{\bf k}}^{2}\sum_{i}\left[\zeta_{i}^{(1)}k_{i}^{4}+\zeta_{i}^{(2)}k_{i}^{2}k_{i+1}^{2}\right],\label{InTE}
\end{equation}
with $\zeta^{(1)}=(25,25,0)$ and $\zeta^{(2)}=(50,64,64)$. In this
case, $\mathring{\delta}({\bf k})$ is present in the entire momentum
space except for the $z$-axis.

The GLS for the $p$-wave $x^{2}-y^{2}$ state in both
$S=1$ ($S=3$) channels can also be described by Eq. (\ref{InTE})
with coefficients $\zeta^{(1)}=(1,1,1)$ and $\zeta^{(2)}=(3,0,0)$
($\zeta^{(1)}=(25/4,25/4,25),\zeta^{(2)}=(103/2,41,41)$). Hence,
the $S=3$ channel demonstrates fully GLSs while the $S=1$ channel
exhibits vanishing $\mathring{\delta}({\bf k})$ along the $[0,0,1]^*$ direction.

Finally, we look at the septet state denoted by $A_{2u}$. In this case, the pairing of electrons with different quantum numbers $m_j$ manifests itself by
\begin{align}
\!\mathring{\delta}({\bf k})\! & \!=\!\!\frac{3\Delta_{{\bf k}}^{2}}{16}\!\Big\{\!\!\sum_{i}\!\!\left[4k_{i}^{6}\!-\!3\left(\!k_{i}^{4}k_{i+1}^{2}\!+\!k_{i}^{4}k_{i+2}^{2}\right)\right]\!\!+\!6k_{x}^{2}k_{y}^{2}k_{z}^{2}\!\Big\}\!,\!\!\!
\end{align}
where $i+2=z$ if $i=x$ and the GLSs are present throughout the momentum space
except for the $[1,1,1]^*$ direction.

\textit{Candidate systems with $T_{d}$ structure}.\textemdash It
is worthwhile to note that the half-Heusler compounds \textit{R}PdBi
have tetrahedral $T_{d}$ symmetry (subgroup of $O_{h}$) without
inversion center. Nevertheless, the formalism of describing the pairing
is the same as for the $O_{h}$ group but different IR labels apply,
\textit{cf.} Table \ref{Table1}. The non-centrosymmetry manifests
itself by an antisymmetric spin-orbit coupling (ASOC) given by \citep{Brydon,InvBreak1}
\begin{equation}
\hat{H}^{\prime}({\bf k})=\delta\sum_{i}k_{i}\left(\hat{J}_{i+1}\hat{J}_{i}\hat{J}_{i+1}-\hat{J}_{i+2}\hat{J}_{i}\hat{J}_{i+2}\right),\label{ASOC}
\end{equation}
where $\delta$ controls the strength of the ASOC and $i\in\{x,y,z\}$.
Projecting $\hat{H}^{\prime}({\bf k})$ onto the intra-band basis, this results in splitting the energy band as $E_{{\bf k}}^{\nu}\rightarrow E_{{\bf k}}^{\nu}\pm|\boldsymbol{g}_{{\bf k}}^{\nu\nu}|$
with $\boldsymbol{g}_{{\bf k}}^{\nu\nu}.\boldsymbol{\sigma}=\hat{V}_{{\bf k}}^{\nu\dagger}\hat{H}^{\prime}({\bf k})\hat{V}_{{\bf k}}^{\nu}$
and $\nu=\pm$, as shown in Figs. \ref{Fig2}(b) and \ref{Fig2}(c)
\citep{cm}. Here, $\boldsymbol{g}_{{\bf k}}^{\nu\nu}=(g_{x}^{\nu\nu},g_{y}^{\nu\nu},g_{z}^{\nu\nu})$
and $\boldsymbol{\sigma}=(\hat{\sigma}_{x},\hat{\sigma}_{y},\hat{\sigma}_{z})$
are momentum dependent ASOC vector and Pauli matrices in the intra-band
basis, respectively. The lack of inversion symmetry allows the pairing
state to be a mixture of even-parity singlet $\hat{\mathcal{H}}_{I}^{0,0,0}({\bf k})$
and odd-parity $p$-wave states \citep{Noncentro}. In this case,
the most stable odd-parity pairing state with the largest transition
temperature may arise when its ${\bf d}$-vector aligns parallel to
the ASOC vector \citep{ArxSeptEx,Agterberg}. Thus, by combining $\hat{H}^{\prime}({\bf k})$
with the Cooper pair symmetrization matrix $\hat{\mathcal{R}}=i\hat{\sigma}_{x}\otimes\hat{\sigma}_{y}$
in the $j=3/2$ representation, we arrive at the septet
pairing state $\hat{\mathcal{H}}_{xyz}^{3,3,1}({\bf k})=\hat{H}^{\prime}({\bf k})\hat{\mathcal{R}}$
\citep{Brydon}. The inter-band crossing of the mixed superconducting
state $\hat{\mathcal{H}}_{I}^{0,0,0}({\bf k})+\hat{\mathcal{H}}_{xyz}^{3,3,1}({\bf k})$
cannot be hybridized by the inversion symmetry breaking ASOC. Therefore,
the emergence of finite-energy superconducting coherence peaks in
the DOS are strong indicators of septet Cooper pairing of electrons with different quantum numbers $m_j$, as shown in Figs.\ \ref{Fig2}(a) and \ref{Fig2}(b). Note
the difference to singlet pairing, where the DOS exhibit
a flat shape away from the Fermi surface, \textit{cf.} Fig.\ \ref{Fig2}(c).
Remarkably, both odd- and even-parity channels of $3z^{2}-r^{2}$
turn into fully GLSs in the presence of ASOC, \textit{cf.} Table \ref{Table1}.
This also partially happens for the $A_{2}$ state and the ($x^{2}-y^{2}$,2,1,1)
state. Therefore, a small value of ASOC even enhances the likelihood of observing GLSs in the
DOS.

To observe the $m_j$ content of the novel pairing at FEE, we propose to apply a perturbative Zeeman field to the system where the states acquire finite magnetization in terms of $m_j$ degrees of freedom due to broken time-reversal symmetry \citep{Supp}. Consequently, the GLSs split into two different pairs of GLSs. Each GLS corresponds to paired  electrons with different magnetic quantum numbers signaled by simultaneous drops in the  $m_j$-resolved DOS.

\textit{Conclusions.}\textemdash We have investigated Cooper
pairing in $j=3/2$ superconductors with cubic point-group symmetry.
The multiband nature of the system with identical bending configuration
allows for observing Cooper pairing away from the Fermi surface in
the weak pairing limit. This manifests itself by a pair of indirect
finite-energy anti-crossings of BdG bands signaling pairing of electrons having different components of total angular momentum. The phenomenon may be experimentally detectable through
tunneling spectroscopy \citep{BulkSpec,STS,Spect,TunSpec} and angle-resolved
photo-emission spectroscopy \citep{ARPES}.

\textit{Note added.}\textemdash During the preparation of this manuscript,
we became aware of a related proposal of inter-band pairing away from
the Fermi surface. This proposal is about the emergence of anapole superconductivity in the presence of competing pairing channels. Hence, the physics is different from ours \citep{Anapole}.

We thank M. Bode, S. J. Choi, P. Eck,  M. V. Hosseini, C. A. Li, G. Sangiovanni  and A. H. Talebi  for fruitful discussions. The work was supported by the DFG (SPP1666 and
SFB1170 ToCoTronics), the Würzburg-Dresden Cluster of Excellence ct.qmat,
EXC2147, Project Id 390858490, and the Elitenetzwerk Bayern Graduate
School on Topological Insulators.

\providecommand{\noopsort}[1]{}\providecommand{\singleletter}[1]{#1}%

  \newpage
  \appendix
\begin{center}
\bf{\large Supplemental Material}
\end{center}
\maketitle
\tableofcontents{}
\section{Folding-down approach \label{App: Folding-down}}

In this section, we show the derivation of the effective Hamiltonian
presented in Eq. (5) of the Letter through the folding down approach
\citep{Fold}. Consider the following Schrödinger equation 
\begin{align}
\left(\begin{array}{cc}
\hat{H}_{11} & \hat{H}_{12}\\
\hat{H}_{21} & \hat{H}_{22}
\end{array}\right)\left(\begin{array}{c}
\hat{\psi}_{A}\\
\hat{\psi}_{B}
\end{array}\right) & =E\left(\begin{array}{c}
\hat{\psi}_{A}\\
\hat{\psi}_{B}
\end{array}\right),
\end{align}
where $\hat{H}_{ij}$ is a $n\times n$ sub-block matrix and $(\hat{\psi}_{A},\hat{\psi}_{B})^{T}$
denotes the eigenvector column with $\hat{\psi}_{A(B)}$ being its
sub-block elements. The above eigenvalue problem reduces to the following
coupled equations 
\begin{align}
\hat{H}_{11}\hat{\psi}_{A}+\hat{H}_{12}\hat{\psi}_{B} & =E\hat{\psi}_{A},\label{App: Fold Eq1}\\
\hat{H}_{21}\hat{\psi}_{A}+\hat{H}_{22}\hat{\psi}_{B} & =E\hat{\psi}_{B}.\label{App: Fold Eq2}
\end{align}
From Eq. (\ref{App: Fold Eq2}), we obtain $\hat{\psi}_{B}$ as 
\begin{align}
\hat{\psi}_{B} & =(E\hat{I}_{n}-\hat{H}_{22})^{-1}\hat{H}_{21}\hat{\psi}_{A},
\end{align}
where $\hat{I}_{n}$ is a $n\times n$ identity matrix. Inserting
$\hat{\psi}_{B}$ into Eq. (\ref{App: Fold Eq1}) results in

\begin{align}
\mathscr{\hat{H}}_{\text{eff}}\ \hat{\psi}_{A} & =E\hat{\psi}_{A},\label{Shrodinger}
\end{align}
where 
\begin{align}
\mathscr{\hat{H}}_{\text{eff}}=\hat{H}_{11}+\hat{H}_{12}(\omega\hat{I}_{n}-\hat{H}_{22})^{-1}\hat{H}_{21}.\label{Eff}
\end{align}
Note that the second term in $\mathscr{\hat{H}}_{\text{eff}}$ has
the same basis as $\hat{H}_{11}$. Moreover, we have taken $E\rightarrow E-\omega+\omega$
where $E-\omega\approx0$ holds in the vicinity of finite-excitation
energies (FEE) $\omega$ and making the left hand side of Eq. (\ref{Shrodinger})
independent of $E$. The effective low-energy pairing can be easily
derived by rearranging the inter-band basis of the sub-block Hamiltonians
into the intra-band basis as well as setting $\omega\approx0$.

\section{Power series expansion} \label{App: Power series expansion}

Equation (4) of the Letter can be represented in sub-block matrix
formalism as
\begin{equation}
\hat{h}_{k}=\left(\begin{array}{cc}
\hat{h}_{11}({\bf k}) & \hat{h}_{12}({\bf k})\\
\hat{h}_{21}({\bf k}) & \hat{h}_{22}({\bf k})
\end{array}\right),\label{subBlock}
\end{equation}
which allows us to employ Eq. (\ref{Eff}). To perform the power series
expansion of $(\omega\hat{I}_{4}-\hat{h}_{22}({\bf k}))^{-1}$, we
follow the subsequent steps. Suppose that we are seeking $(\hat{A}+\hat{B})^{-1}$
where $\hat{A}$ and $\hat{B}$ are invertible Hermitian matrices
of dimension $n$. Consider the following identity
\begin{equation}
(\hat{A}+\hat{B})^{-1}=\hat{A}^{-1}(\hat{I}_{n}+\hat{B}\hat{A}^{-1})^{-1}.\label{App: powerseries}
\end{equation}
A power series expansion of the right hand side of Eq. (\ref{App: powerseries})
reads
\begin{align}
(\hat{I}_{n}\!+\!\hat{B}\hat{A}^{-1})^{-1} & \!=\!\sum_{n=0}^{\infty}(-\hat{B}\hat{A}^{-1})^{n}\!\nonumber \\
 & \!=\!\hat{I}_{n}\!-\!\hat{B}\hat{A}^{-1}\!+\!\mathcal{O}(BA^{-1})^{2}\!.
\end{align}
Inserting the above expansion into the right hand side of Eq. (\ref{App: powerseries}),
this results in 
\begin{equation}
(\hat{A}+\hat{B})^{-1}\approx\hat{A}^{-1}-\hat{A}^{-1}\hat{B}\hat{A}^{-1}.\label{App:InversePowerS}
\end{equation}
Now, expressing $\omega\hat{I}_{4}-\hat{h}_{22}(k)$ in terms of normal
$\mathscr{\hat{E}}_{{\bf k}}$ and pairing $\hat{\triangle}_{{\bf k}}$
parts, this results in the following form 
\begin{equation}
[\omega\hat{I}-\hat{h}_{22}({\bf k})]=\mathscr{\hat{E}}_{{\bf k}}-\hat{\triangle}_{{\bf k}},
\end{equation}
with
\begin{align}
\mathscr{\hat{E}}_{{\bf k}} & \!\equiv\!\left(\!\!\begin{array}{cc}
(\omega\!-\!E_{{\bf k}}^{-})\hat{\sigma}_{0} & 0\\
0 & (\omega\!+\!E_{{\bf k}}^{+})\hat{\sigma}_{0}
\end{array}\!\!\right),\nonumber \\
\hat{\triangle}_{{\bf k}} & \!\equiv\!\!\left(\!\!\begin{array}{cc}
0 & \hat{\Delta}_{{\bf k}}^{-+}\\
(\hat{\Delta}_{{\bf k}}^{-+})^{\dagger} & 0
\end{array}\!\!\right)\!.
\end{align}
Using Eq. (\ref{App:InversePowerS}), we arrive immediately at 
\begin{align}
[\omega\hat{I}-\hat{h}_{22}({\bf k})]^{-1} & \approx\mathscr{\hat{E}}_{{\bf k}}^{-1}+\mathscr{\hat{E}}_{{\bf k}}^{-1}\hat{\triangle}_{{\bf k}}\mathscr{\hat{E}}_{{\bf k}}^{-1},
\end{align}
where we have assumed that $\hat{\triangle}_{{\bf k}}$ in the vicinity
of $\tilde{\boldsymbol{k}}$ is small. Sandwiching the above term
between off-diagonal blocks of Eq. (\ref{subBlock}), we arrive at
\begin{equation}
\hat{\varLambda}({\bf k})=\hat{h}_{12}({\bf k})[\omega\hat{I}-\hat{h}_{22}({\bf k})]^{-1}\hat{h}_{21}({\bf k}),
\end{equation}
which reads explicitly
\begin{align}
\!\!\hat{\varLambda}({\bf k})\!\!=\!\!\!\left(\!\!\!\begin{array}{cc}
\!\!\!\frac{1}{(\omega+E_{{\bf k}}^{+})}\Delta_{{\bf k}}^{\!++}(\!\Delta_{{\bf k}}^{\!++}\!)^{\dagger} & \frac{1}{\varepsilon_{{\bf k}}}\Delta_{{\bf k}}^{\!++}(\!\Delta_{{\bf k}}^{\!-+}\!)^{\dagger}\Delta_{{\bf k}}^{\!--}\\
\frac{1}{\varepsilon_{{\bf k}}}[\Delta_{{\bf k}}^{\!++}(\!\Delta_{{\bf k}}^{\!-+}\!)^{\dagger}\Delta_{{\bf k}}^{\!--}]^{\dagger} & \frac{1}{(\omega-E_{{\bf k}}^{-})}(\!\Delta_{{\bf k}}^{\!--}\!)^{\dagger}\Delta_{{\bf k}}^{\!--}
\end{array}\!\!\!\right)\!\!,\!\!\!\!
\end{align}
 where $\varepsilon_{{\bf k}}=(\omega+E_{{\bf k}}^{+})(\omega-E_{{\bf k}}^{-})$.
Finally, adding the above term to $\hat{h}_{11}({\bf k})$, this results
in Eq. (5) of the Letter. In our choice of time-reversal symmetric
pairing, the identity $(\!\Delta_{{\bf k}}^{\!\nu\nu}\!)^{\dagger}\Delta_{{\bf k}}^{\!\nu\nu}=\!\Delta_{{\bf k}}^{\!\nu\nu}(\!\Delta_{{\bf k}}^{\!\nu\nu})^{\dagger}$
with $\nu\in\{+,-\}$, holds true since $(\!\Delta_{{\bf k}}^{\!\nu\nu}\!)^{\dagger}\Delta_{{\bf k}}^{\!\nu\nu}$
is proportional to the identity matrix.

\section{Band basis formalism\label{App: Band basis}}

By exact diagonalization of Eq. (1) of the main Letter for the case
of SO(3) symmetry, i.e., $\gamma=\beta$, we obtain the eigenvector
matrix $\hat{V}_{{\bf k}}^{\pm}$ corresponding to the two-fold degenerate
eigenvalues $E_{{\bf k}}^{\pm}$ as 
\begin{align}
\hat{V}_{{\bf k}}^{+} & =\Gamma_{{\bf k}}^{+}\left(\begin{array}{cc}
2k_{z}k_{-}/k_{+}^{2} & k_{-}/k_{+}\\
\sqrt{3}k_{-}/k_{+} & 0\\
0 & \sqrt{3}\\
1 & -2k_{z}/k_{-}
\end{array}\right),
\end{align}
and 
\begin{align}
\hat{V}_{{\bf k}}^{-} & \!\!=\!\!\Gamma_{{\bf k}}^{-}\!\!\left(\begin{array}{cc}
2\sqrt{3}k_{z}k_{-}(k_{x}^{2}+k_{y}^{2})/k_{+}^{2} & -\sqrt{3}k_{-}^{2}\\
-(k+3k_{z}^{2})(k_{x}^{2}+k_{y}^{2})/k_{+}^{2} & 0\\
0 & k+3k_{z}^{2}\\
\sqrt{3}(k_{x}^{2}+k_{y}^{2}) & 2\sqrt{3}k_{z}k_{+}
\end{array}\right),
\end{align}
where $\Gamma_{{\bf k}}^{+}=\sqrt{k_{x}^{2}+k_{y}^{2}}/2k$, $\Gamma_{{\bf k}}^{-}=(2k\sqrt{k+3k_{z}^{2}})^{-1}$,
and $k_{\pm}=k_{x}\pm ik_{y}$. Note that the eigenvector matrix $\hat{V}=(\hat{V}_{{\bf k}}^{+},\hat{V}_{{\bf k}}^{-})$
is orthonormal (derived by the Gram\textendash Schmidt method) satisfying
$\hat{V}^{\dagger}\hat{V}=\hat{I}_{4}$. The relation between the
band basis (pseudo-spin) operator $\hat{f}_{{\bf k}}^{\pm}$ and the
fermionic operator is given by 
\begin{equation}
\hat{f}_{{\bf k}}^{\pm\dagger}=\hat{c}_{{\bf k}}^{\dagger}\hat{V}_{{\bf k}}^{\pm}.
\end{equation}
The band basis operators should transform under time-reversal and
inversion operations as usual fermionic operators. To construct such
correspondence, we act with the anti-unitary time-reversal operation
$\Theta$ on the fermionic basis $c_{{\bf k},m}$ in the usual way
\begin{equation}
\Theta c_{{\bf k},m_{j}}=(-1)^{j+m_{j}}c_{-{\bf k},-m_{j}},\label{App:time-revFerm}
\end{equation}
 where the electron annihilation operator acquires a phase in addition
to a sign change of momentum and magnetic quantum number $m_{j}$.
To perform the above operation on the basis of the normal state Hamiltonian
$\hat{c}_{{\bf k},m}=(c_{{\bf k},3/2},c_{{\bf k},1/2},c_{{\bf k},-1/2},c_{{\bf k},-3/2})^{T}$,
we introduce the matrix representation of the time-reversal operator
in $j=3/2$ basis as $\hat{\Theta}=\hat{\mathcal{T}}\mathcal{K}$
with $\hat{\mathcal{T}}=i\hat{\sigma}_{x}\otimes\hat{\sigma}_{y}$
and $\mathcal{K}$ being the unitary part of the time-reversal operator
and the complex conjugate operator, respectively. Therefore, the time-reversal
transformation of the normal-state basis takes the form
\begin{equation}
\hat{\mathcal{T}}^{-1}\hat{c}_{{\bf k}}=\hat{c}_{-{\bf k}}^{\prime},\label{App:time-revnormal}
\end{equation}
where $\hat{c}_{-{\bf k}}^{\prime}$ is the column of time-reversed
fermionic operators given by 
\begin{align}
\hat{c}_{-{\bf k}}^{\prime} & =(-c_{-{\bf k},-3/2},c_{-{\bf k},-1/2},-c_{-{\bf k},1/2},c_{-{\bf k},3/2})^{T}.
\end{align}
Note that $\hat{\mathcal{T}}$ fulfills the property $\hat{\mathcal{T}}^{2}=-\hat{I}_{4}$.
Also, the pseudo-spin operator $f_{{\bf k},\uparrow\downarrow}$ under
time-reversal operation obeys Eq. (\ref{App:time-revFerm}) as 
\begin{equation}
\hat{\mathscr{T}}^{-1}\hat{f}_{{\bf k}}^{\pm}=\hat{f}_{-{\bf k}}^{\prime\pm},\label{App:time-revpseudo}
\end{equation}
 with $\hat{\mathscr{T}}=i\hat{\sigma}_{y}$ denoting the $2\times2$
matrix representation of the unitary part of the time-reversal operator
in pseudo-spin$-1/2$ basis, and
\begin{align}
\hat{f}_{-{\bf k}}^{\prime\pm} & =(-f_{-{\bf k},\downarrow}^{\pm},f_{-{\bf k},\uparrow}^{\pm}).
\end{align}
Note that we have taken into account the pseudo-spin index as effective
spin-1/2 index. Inserting Eqs. (\ref{App:time-revpseudo}) into Eq.
(\ref{App:time-revnormal}), this results in 
\begin{align}
V_{-{\bf k}}^{\prime}=-\hat{\mathcal{T}}V_{{\bf k}}^{*}(i\sigma_{y}),
\end{align}
with $V_{-{\bf k}}^{\prime}$ being the time-reversed matrix of eigenvectors.
Moreover, the band basis operators satisfy inversion symmetry as $\hat{V}_{{\bf k}}^{\pm}=\hat{V}_{-{\bf k}}^{\pm}$.

\section{Angular-momentum-resolved density of states}

In this section, we derive the angular-momentum-resolved density of
states (DOS) in the BdG formalism for $j=3/2$ superconductors. Consider
the Luttinger model in the superconducting phase described by the
BdG Hamiltonian in Eq. (3) of the Letter. Diagonalizing the superconducting
Hamiltonian $H=\sum_{{\bf k}}\hat{\psi}_{{\bf k}}^{\dagger}\hat{H}_{\text{BdG}}({\bf k})\hat{\psi}_{{\bf k}}$,
this results in 
\begin{align}
H & =\sum_{{\bf k}}\hat{a}_{{\bf k}}^{\dagger}\ \hat{\mathbb{E}}_{{\bf k}}\ \hat{a}_{{\bf k}},
\end{align}
with the basis

\begin{equation}
\hat{a}_{{\bf k}}^{\dagger}\!=\!(a_{{\bf k},1}^{\dagger},a_{{\bf k},2}^{\dagger},a_{{\bf k},3}^{\dagger},a_{{\bf k},4}^{\dagger},a_{{\bf k},5}^{\dagger},a_{{\bf k},6}^{\dagger},a_{{\bf k},7}^{\dagger},a_{{\bf k},8}^{\dagger}),\!\!\!
\end{equation}
where $a_{{\bf k},i}^{\dagger}$ is the creation operator for BdG
quasi-particles of the $i$th excitation band and $\hat{\mathbb{E}}_{{\bf k}}=\!\hat{\mathcal{U}}_{{\bf k}}^{\dagger}\!\hat{H}_{\text{BdG}}({\bf k})\hat{\mathcal{U}}_{{\bf k}}$
is a $8\times8$ diagonal matrix of eigenvalues, with $\hat{\mathcal{U}}_{{\bf k}}$
the matrix of eigenspinors, given by 

\begin{align}
\!\!\hat{\mathbb{E}}_{{\bf k}}\!\!=\!\text{diag}\big( & \!E_{{\bf k},1},\!E_{{\bf k},2},\!E_{{\bf k},3},\!E_{{\bf k},4},\!E_{{\bf k},5},\!E_{{\bf k},6},\!E_{{\bf k},7},\!E_{{\bf k},8}\!\big).\!\!\!
\end{align}
 The relationship between the basis of the Hamiltonian and its band
basis (eigenbasis) is given by $\hat{a}_{{\bf k}}^{\dagger}=\hat{\psi}_{{\bf k}}^{\dagger}\hat{\mathcal{U}}_{{\bf k}}$
with
\begin{equation}
\hat{\mathcal{U}}_{{\bf k}}\!=\!(\hat{\Phi}_{{\bf k},1},\hat{\Phi}_{{\bf k},2},\hat{\Phi}_{{\bf k},3},\hat{\Phi}_{{\bf k},4},\hat{\Phi}_{{\bf k},5},\hat{\Phi}_{{\bf k},6},\hat{\Phi}_{{\bf k},7},\hat{\Phi}_{{\bf k},8}),\!\!
\end{equation}
where $\hat{\Phi}_{{\bf k},i}$ is the eigenspinor corresponding to
$E_{{\bf k},i}$ with $i=\{1,2,3,4,5,6,7,8\}$ being the band indices.
Each $\hat{\Phi}_{{\bf k},i}$ is comprised of electron ($e$) and
hole ($h$) probability weights denoted by $\hat{\Phi}_{{\bf k},i}=(\hat{\Phi}_{{\bf k},i}^{e},\hat{\Phi}_{{\bf k},i}^{h})^{T}$.
The electron (hole) components are given by

\begin{equation}
\hat{\Phi}_{{\bf k},i}^{e(h)}=(u_{{\bf k},i,\frac{3}{2}}^{e(h)},u_{{\bf k},i,\frac{1}{2}}^{e(h)},u_{{\bf k},i,-\frac{1}{2}}^{e(h)},u_{{\bf k},i,-\frac{3}{2}}^{e(h)}).\label{elec-hole comp}
\end{equation}
In Eq. ($\text{\ref{elec-hole comp}}$), the components are labeled
by the magnetic quantum number $m_{j}=\pm3/2,\pm1/2$ due to the choice
of basis. According to the above description, the angular-momentum-resolved
DOS in the BdG formalism takes the form

\begin{equation}
N_{m_{j}}\!(E)\!=\!\!\sum_{i=1}^{N}\!\sum_{{\bf k}}\delta(E-E_{{\bf k},i})\big(|u_{{\bf k},m_{j}}^{e}|^{2}+|u_{{\bf k},m_{j}}^{h}|^{2}\big),\!\label{mj resolved}
\end{equation}
with $N=8$ being the total number of excitation energy bands. The
total superconducting DOS can be derived by taking into account the
contribution of all $m_{j}$ components of the DOS given by

\begin{align}
N(E)\!=\!\!\sum_{m_{j}}\!N_{m_{j}}\!(E)\!=\!\!\sum_{i=1}^{N}\!\sum_{{\bf k}}\delta(E-E_{{\bf k}}^{i}),\label{Total Dos}
\end{align}
where $m_{j}\in\{3/2,1/2,-1/2,-3/2\}.$ Eq. ($\text{\ref{Total Dos}}$)
is simplified due to normalization condition

\begin{equation}
\sum_{m_{j}}\big(|u_{{\bf k},m_{j}}^{e}|^{2}+|u_{{\bf k},m_{j}}^{h}|^{2}\big)=1.
\end{equation}

\section{Proposal to detect \lowercase{$j$}$=3/2$ pairing}

To address the observability of the $m_{j}$ structure of $j=3/2$
Cooper pairing, we propose to investigate the $m_{j}$-resolved DOS
in the presence of a perturbative Zeeman field. The corresponding
term in the Hamiltonian is given by $\hat{B}=\boldsymbol{\mathcal{M}}\cdot\hat{\boldsymbol{J}}$
with $\boldsymbol{\mathcal{M}}=(\mathcal{M}_{x},\mathcal{M}_{y},\mathcal{M}_{z})$
being the Zeeman vector and $\hat{\boldsymbol{J}}=(\hat{J}{}_{x},\hat{J}_{y},\hat{J}_{z})$
defining the vector of $j=3/2$ matrices. The doubly degenerate energy
bands have zero net magnetization throughout momentum space due to
the combination of time-reversal and inversion symmetries. In the
presence of the Zeeman field, the energy bands incorporating states
with different magnetic quantum number $m_{j}$ are split as a consequence
of time reversal symmetry breaking and they acquire finite magnetization.

\begin{figure}
\begin{centering}
\includegraphics[scale=0.5]{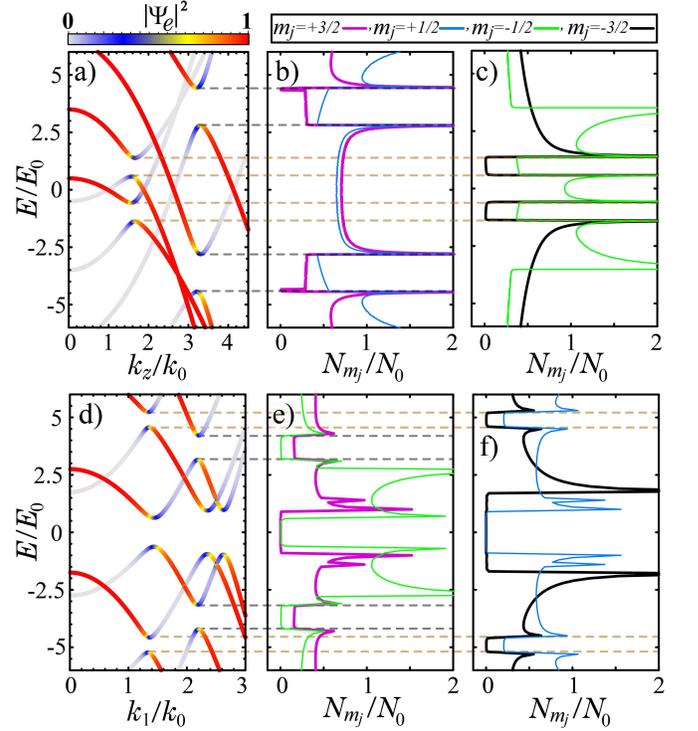}
\par\end{centering}
\caption{\label{Sup-Fig2} Superconducting spectra of the p-wave septet pairing
in (a) $[0,0,1]$ and (d) $[1,1,0]$ (i.e., $k_{x}=k_{y}=k_{1}$ and
$k_{z}=0$) directions for parameters (a) $(\Delta/E_{0}a,\delta/E_{0}a,m_{z}/E_{0})=(3,0,3)$
and (d) $(\Delta/E_{0}a,\delta/E_{0}a,m_{z}/E_{0})=(4,0.5,4.5)$,
respectively. The color denotes the probability of electronic states
$|\Psi_{e}|^{2}$ in panels (a) and (d). The $m_{j}-$resolved DOS
of panel (a) {[}(d){]} is depicted in panels (b) and (c) {[}(e) and
(f){]}. Other parameters are $\mu/E_{0}=-5$, $k_{0}=0.1a^{-1}$,
$E_{0}=10^{-2}|\alpha|a^{-2}$, $\beta=0.2|\alpha|$, $\gamma=-0.05|\alpha|$,
$\alpha=-20$, and $N_{0}=10^{5}$.}
\end{figure}

Without loss of generality, we choose the magnetic field to point
in $z$ direction, i.e., $\boldsymbol{\mathcal{M}}=(0,0,\mathcal{M}_{z})$.
We focus on the p-wave septet pairing channel since this is the most
energetically favorable instability channel in half-Heusler materials
with $T_{d}$ crystalline structure \citep{ArxSeptEx}. For a finite
value of $\mathcal{M}_{z}$, the spectrum in Fig. 1(c) of the Letter
is re-plotted in Fig. $\text{\ref{Sup-Fig2}}$(a) below in the presence
of cubic anisotropy. The $m_{j}$-resolved DOS, according to Eq. ($\text{\ref{mj resolved}}$),
is depicted in Fig. $\text{\ref{Sup-Fig2}}$(b) and (c). The solid
lines with magenta, light blue, light green and black colors illustrate
$N_{3/2}(E)$, $N_{1/2}(E)$, $N_{-1/2}(E)$ and $N_{-3/2}(E)$, respectively.

\begin{figure}
\centering{}\includegraphics[scale=0.48]{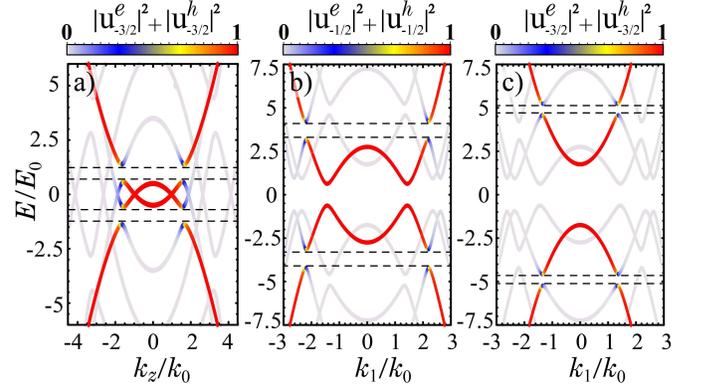}\caption{\label{Fig3-JreslvSpec}Angular-momentum-resolved BdG spectra of the
p-wave septet pairing in (a) $[0,0,1]$ and (b,c) $[1,1,0]$ (i.e.,
$k_{x}=k_{y}=k_{1}$ and $k_{z}=0$) directions. The color bar denotes
the probability weights of quasi-particles with (a) $m_{j}=-3/2$,
(b) $m_{j}=-1/2$ and (c) $m_{j}=-3/2$ magnetic quantum number. Other
parameters are the same as those in Fig.\ $\text{\ref{Sup-Fig2}}$.}
\end{figure}

Fig. $\text{\ref{Sup-Fig2}}$(a) shows that the nodal behavior at
low-energy is remained intact while a pair of GLSs, shown in Fig.
1(c) of the Letter, are split into two pairs of GLSs due to violation
of time-reversal symmetry. We point out that the larger GLSs happen
within the excitation energy ranges $E/E_{0}\in[2.8,4.4]$ and $E/E_{0}\in[-4.4,-2.8]$.
Interestingly, in this energy range, the simultaneous abrupt drops
of the $m_{j}-$resolved DOS in Fig. $\text{\ref{Sup-Fig2}}$(b) signal
hybridization of electrons with different quantum numbers, i.e., $m_{j}=3/2$
with $m_{j}=1/2$. To visualize it, we connect the GLSs energy range
in Fig. $\text{\ref{Sup-Fig2}}$(a) to the energy range of coherence
peaks in Fig. $\text{\ref{Sup-Fig2}}$(b) with gray dashed lines.
In addition, the smaller GLSs appearing in the energy ranges $E/E_{0}\in[0.55,1.35]$
and $E/E_{0}\in[-1.35,-0.55]$ stem from superconducting hybridization
of $m_{j}=-3/2$ and $m_{j}=-1/2$ states. This is signaled by simultaneous
drops of $N_{-3/2}(E)$ and $N_{-1/2}(E)$ in these energy regions
(light brown dashed lines) as shown in Fig. $\text{\ref{Sup-Fig2}}$(c).
Surprisingly, the DOS of $m_{j}=-3/2$ states completely vanishes
at FEE. This is because the $m_{j}=-3/2$ electron band is located
below the hybridization energy $E(\tilde{k})$ where pairing occur
with $m_{j}=-1/2$ states. In this case, there are no states within
the GLS excitation energies as illustrated in Fig. $\text{\ref{Fig3-JreslvSpec}}$(a)
where the dashed lines mark the paired area. This converts the superconducting
GLS into a \textit{full gap} for a particular choice of $m_{j}$ at
finite-energies despite of having multiband structure. The DOS $N_{-3/2}(E)$
in Fig. $\text{\ref{Sup-Fig2}}$(c) corresponds to the $m_{j}=-3/2$
resolved spectrum in Fig. $\text{\ref{Fig3-JreslvSpec}}$(a). Note
also that the $m_{j}$-resolved DOS in Fig. $\text{\ref{Sup-Fig2}}$(b)
and (c) show finite values at $E=0$ due to the nodal behavior.

The observation of $m_{j}-$resolved DOS is challenging, but keeping
in mind to role of multiband systems in modern quantum materials.
In principle, it can be accomplished in a similar way as spin-resolved
spectroscopy. We need a spectrometer (e.g. based on scanning tunneling
spectroscopy) that is able to distinguish electrons with different
magnetic quantum numbers.
We may ask about the reason behind the superconducting hybridization
of states with \textit{equal sign} of $m_{j}$ in the aforementioned
example. The reason is rooted in the anisotropy of the instability
channel in momentum space, in which the paired states with different
$m_{j}$ are affected by the wave vector. Due to the absence of ASOC,
$j_{z}$ is conserved along the z-axis and  it is instructive to
look at the second quantization representation of the pairing channel
given by Eq. (\textbf{$\ref{xyz}$}) below as $H_{xyz}^{3,3,1}\!=\!\sum_{k_{z}}\hat{c}_{k_{z}}^{\dagger}\hat{\mathcal{H}}_{xyz}^{3,3,1}(k_{z})\hat{c}_{-k_{z}}^{\dagger T}$
which can be written as

\begin{equation}
H_{xyz}^{\negthinspace(\negthinspace3,3,1\negthinspace)}\!\!=\!\!\frac{\sqrt{3}}{2}\Delta\!\!\sum_{\mathrm{k}_{z}}\!\mathrm{k}_{z}\big(c_{\frac{3}{2}}^{\dagger}c_{\frac{1}{2}}^{\dagger}\!\!+c_{\frac{1}{2}}^{\dagger}c_{\frac{3}{2}}^{\dagger}\!\!+c_{\!\!-\frac{3}{2}}^{\dagger}c_{\!\!-\frac{1}{2}}^{\dagger}\!\!+c_{\!\!-\frac{1}{2}}^{\dagger}c_{\!\!-\frac{3}{2}}^{\dagger}\!\big)\!\!+\!h.c.,\!\label{BandBasisAlongZ}
\end{equation}
where the momentum dependency of the operators are dropped for ease
of notation. According to Eq. ($\text{\ref{BandBasisAlongZ}}$), we
can realize that the pair operators with different $m_{j}$ and \textit{equal}
signs remain finite along the z-direction due to the anisotropic form
of the pairing channel. Importantly, the larger (smaller) GLSs in
Fig. $\text{\ref{Sup-Fig2}}$(b) {[}(c){]} correspond to $c_{3/2}^{\dagger}c_{1/2}^{\dagger}+c_{1/2}^{\dagger}c_{3/2}^{\dagger}$
($c_{-1/2}^{\dagger}c_{-3/2}^{\dagger}+c_{-3/2}^{\dagger}c_{-1/2}^{\dagger}$)
pairing operators. Note that the presence of nodal degeneracies at
low energies are due to the absence of pairing between states with
identical $m_{j}$.

It is important to mention that the resolution of superconducting
coherence peaks in quantum numbers $m_{j}$ is restricted neither
to the $[0,0,1]$ direction nor the conservation of $j_{z}$. To show
this, we focus on the $C_{2}^{\prime}$ axis, i.e., $[1,1,0]$ direction,
where the conservation of $j_{z}$ is violated due to the presence
of symmetric and antisymmetric spin-orbit coupling. The spectrum,
in the presence of a perpendicular magnetic field $m_{z}=4.5E_{0}$
and a small value of ASOC, is calculated in Fig. $\text{\ref{Sup-Fig2}}$(d).
The split GLSs with larger partial band gap correspond to pairing
of $m_{j}=3/2$ and $m_{j}=-1/2$ states according to the coincident
drops of $N_{m_{j}}(E)$ in Fig. $\text{\ref{Sup-Fig2}}$(e). The
corresponding energy range is enclosed by gray dashed lines. Remarkably,
$N_{-1/2}(E)$ vanishes completely at FEE owing to the fact that the
$m_{j}=-1/2$ electron band is pushed below the superconducting energy
$E(\tilde{k})$ as a consequence of interplay between magnetic field
and Fermi energy. The corresponding $m_{j}=-1/2$ resolved spectra
having zero DOS character (Fig. $\text{\ref{Sup-Fig2}}$(e)) is plotted
in Fig. $\text{\ref{Fig3-JreslvSpec}}$(b) where the full superconducting
gap at FEE is illustrated between the dashed lines. Moreover, the
smaller GLSs in Fig. $\text{\ref{Sup-Fig2}}$(d) are composed by pairing
of $m_{j}=-3/2$ and $m_{j}=1/2$ states according to Fig. $\text{\ref{Sup-Fig2}}$(f)
with entirely vanishing $N_{-3/2}(E)$ at FEE. The reason of $N_{-3/2}(E)$
disappearance is the same as we explained earlier and the $m_{j}=-3/2$
resolved excitations are depicted in Fig. $\text{\ref{Fig3-JreslvSpec}}$(c).

Interestingly, we find that novel Cooper pairs possessing larger magnetic
quantum numbers exhibit larger GLSs in the presence of magnetic fields.
To show this, we identify the total magnetic quantum number of a \textit{local
}finite-energy Cooper pair. We should represent the two-particle state
$|j_{1},j_{2};m_{j_{1}},m_{j_{2}}\rangle\equiv|m_{j_{1}},m_{j_{2}}\rangle$
into a local pair state $|\mathcal{J},m_{\mathcal{J}}\rangle$ given
by

\begin{equation}
\!|m_{j_{1}},\!m_{j_{2}}\!\rangle\!=\!\!\sum_{\mathcal{J},m_{\mathcal{J}}}\!\langle\mathcal{J}\!,m_{\mathcal{J}}|m_{j_{1}}\!,m_{j_{2}}\!\rangle|\mathcal{J}\!,\!m_{\mathcal{J}}\!\rangle,\!\!\label{Local_J_CooperPair}
\end{equation}
where $\langle\mathcal{J},m_{\mathcal{J}}|m_{j_{1}},m_{j_{2}}\rangle$
denotes the Clebsch-Gordan coefficient (CGC), $j_{1}=j_{2}=3/2$ due
to the high-angular momenta nature of the $\Gamma_{8}$ bands, and
$\mathcal{J}$ indicates the total angular momentum of a local Cooper
pair with relative magnetic quantum number $m_{\mathcal{J}}$. The
``local'' term points to those energy bands with distinct indices
contributing to pairing among all the energy bands. For instance,
according to Figs. $\text{\ref{Sup-Fig2}}$(b) and (c), the larger
GLSs in Fig. $\text{\ref{Sup-Fig2}}$(b) correspond to Cooper pairing
formed by single-particle state with quantum numbers $m_{j}=3/2$
and $m_{j}=1/2$ as

\begin{equation}
|\frac{3}{2},\frac{1}{2}\rangle+|\frac{1}{2},\frac{3}{2}\rangle\propto|\mathcal{J}=3,m_{\mathcal{J}}=2\rangle.\label{LocalLargJ3}
\end{equation}
 Likewise, the smaller GLSs in Fig. $\text{\ref{Sup-Fig2}}$(c) correspond
to Cooper pairing with quantum numbers

\begin{equation}
|-\frac{3}{2},-\frac{1}{2}\rangle+|-\frac{1}{2},-\frac{3}{2}\rangle\propto|\mathcal{J}=3,m_{\mathcal{J}}=-2\rangle.\label{LocalSmallJ3}
\end{equation}
From Eqs. ($\text{\ref{LocalLargJ3}}$) and ($\text{\ref{LocalSmallJ3}}$),
we confirm that the exotic pairings have septet total angular momentum
$\mathcal{J}=3$ with $m_{\mathcal{J}}=2$ and $m_{\mathcal{J}}=-2$
magnetic quantum numbers signaled by the larger and smaller GLSs,
respectively.

It is worthwhile to mention that the above results do not rely on
the model parameters. They are not even restricted to special directions
in momentum space. The physics remains valid within the entire momentum
space and applies to other pairing channels.

\section{Spectrum of $j=3/2$ pairing at finite-excitation energies}

In this section, we present two examples to elucidate the effective
non-BdG two-band model given in Eq. (7) of the Letter. The model captures
superconducting spectrum close to GLS at FEE. We start by focusing
on the p-wave septet pairing to calculate $\text{Tr}(\hat{\varepsilon}_{{\bf k}}^{\nu\nu})=\frac{1}{2}\text{Tr}\big(\hat{\Delta}_{{\bf k}}^{\nu\nu}(\hat{\Delta}_{{\bf k}}^{\nu\nu})^{\dagger}\big)/(\omega+\nu E_{{\bf k}}^{\nu})$
for doubly degenerate bands given by indices $\nu\in\{+,-\}$. For
simplicity, we assume $\gamma=\beta$ resulting in the normal state
spectra $E_{{\bf k}}^{+}=(\alpha+9\beta/4)k^{2}-\mu$ and $E_{{\bf k}}^{-}=(\alpha+\beta/4)k^{2}-\mu$.
In this case, we have \begin{widetext}

\begin{align}
\frac{1}{2}\text{Tr} & \big(\hat{\Delta}_{{\bf k}}^{++}(\hat{\Delta}_{{\bf k}}^{++})^{\dagger}\big)=81\Delta_{{\bf k}}^{2}(k_{x}^{2}+k_{y}^{2})(k_{x}^{2}+k_{z}^{2})(k_{y}^{2}+k_{z}^{2}),\label{delPP}\\
\frac{1}{2}\text{Tr} & \big(\hat{\Delta}_{{\bf k}}^{--}(\hat{\Delta}_{{\bf k}}^{--})^{\dagger}\big)=9\Delta_{{\bf k}}^{2}\Big(k_{x}^{4}(k_{y}^{2}+k_{z}^{2})+k_{y}^{4}(k_{x}^{2}+k_{z}^{2})+k_{z}^{4}(k_{x}^{2}+k_{y}^{2})-6k_{x}^{2}k_{y}^{2}k_{z}^{2}\Big),\label{delMM}\\
\mathring{\delta}({\bf k}) & =3\Delta_{{\bf k}}^{2}\Big(4(k_{x}^{6}+k_{y}^{6}+k_{z}^{6})-3\big[k_{x}^{4}(k_{y}^{2}+k_{z}^{2})+k_{y}^{4}(k_{x}^{2}+k_{z}^{2})+k_{z}^{4}(k_{x}^{2}+k_{y}^{2})-2k_{y}^{2}k_{z}^{2}k_{x}^{2}\big]\Big),
\end{align}
\end{widetext}with $\Delta_{{\bf k}}=(\Delta/4k)^{2}.$ Note that
Eq. ($\text{\ref{delPP}}$) (Eq. ($\text{\ref{delMM}}$)) indicates
the magnitude of low-energy Cooper pairing composed by two intra-band
electrons having identical magnetic quantum numbers $m_{j}=3/2$ ($m_{j}=1/2$).

\begin{figure}
\begin{centering}
\includegraphics[scale=0.5]{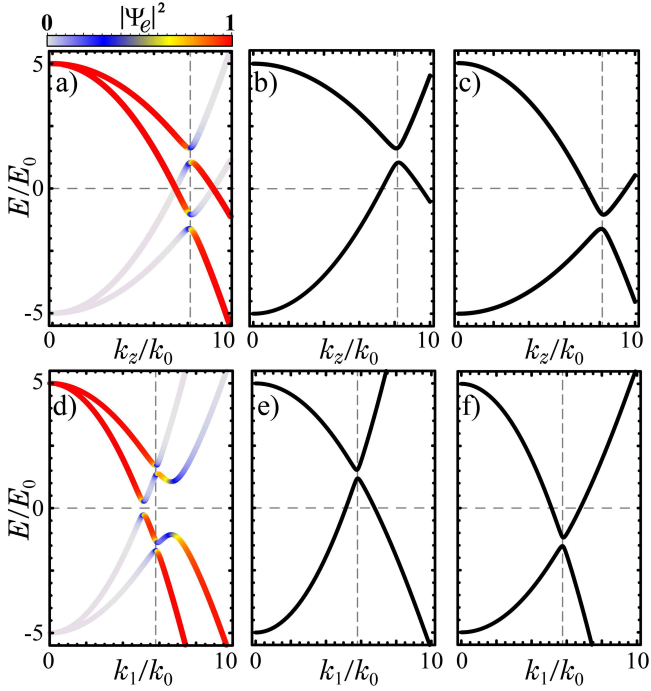}
\par\end{centering}
\caption{\label{Sup-Fig1} BdG spectra (a) and (d) resemble to those in Figs.
1(c) and 1(d) in the Letter, respectively. The right panels (b) and
(c) {[}(e) and (f){]} denote the non-BdG effective spectra calculated
by Eq. (7) of the Letter to capture FEE pairing. The pairing strength
is (a) $\Delta/E_{0}a=4.15$ (b) $\Delta/E_{0}a=10$. Other parameters
are $\beta=0.2|\alpha|$, $\gamma=\beta$, $\mu/E_{0}=-5$, $k_{0}=10^{-2}a^{-1}$,
$E_{0}=10^{-3}|\alpha|a^{-2}$, and $\alpha=-20$. The vertical dashed
lines indicate the exotic pairing momentum for (a) $\tilde{k}=2\sqrt{\mu/(4\alpha+5\beta)}$
and (d) $\tilde{k}=\sqrt{2\mu/(4\alpha+5\beta)}$. The horizontal
line denotes the Fermi surface.}
\end{figure}

We derive the non-BdG two-band spectra corresponding to Figs.1(c)
and (d) of the Letter. The low-energy pairing vanishes along $[0,0,1]$
direction leading to $\text{Tr}(\hat{\varepsilon}_{{\bf k}}^{++})=\text{Tr}(\hat{\varepsilon}_{{\bf k}}^{--})=0$.
Also, the finite-energy pairing becomes $\mathring{\delta}({\bf k})=3\Delta^{2}k_{z}^{2}/4$.
Consequently, the FEE spectrum takes the form

\begin{equation}
\mathcal{E}_{\pm}(k_{z})\!=\!\beta k_{z}^{2}\!\pm\!\!\sqrt{\frac{\vartheta^{2}}{16}k_{z}^{4}\!-\!\frac{\vartheta\mu}{2}k_{z}^{2}\!+\!\mu^{2}\!+\!\frac{3}{4}k_{z}^{2}\Delta^{2}},\label{FEE1}
\end{equation}
with $\vartheta=4\alpha+5\beta.$ Eq. ($\text{\ref{FEE1}}$) denotes
the exotic pairing occurring for positive excitation energies as shown
in Fig. $\text{\ref{Sup-Fig1}}$(b). The vertical dashed line indicates
$k_{z}=\tilde{k}=2\sqrt{\mu/(4\alpha+5\beta)}$ where pairing between
$m_{j}=3/2$ and $m_{j}=1/2$ electrons happen. We can witness in
Fig. $\text{\ref{Sup-Fig1}}$ (b) that the GLS happening at $k_{z}=\tilde{k}$
is in a excellent agreement with the spectrum derived directly by
the BdG Hamiltonian in Fig. $\text{\ref{Sup-Fig1}}$(a). Likewise,
the GLS below the Fermi level is plotted in Fig. $\text{\ref{Sup-Fig1}}$(c).
In both panels (b) and (c), the particle-hole symmetry is broken due
to presence of non-identical diagonal entries, \textit{cf.} Eq. (5)
of the Letter.

Furthermore, the low-energy pairing along the $[1,1,0]$ direction
in both energy bands is finite leading to two superconducting gaps
at the Fermi surface as shown in Fig. $\text{\ref{Sup-Fig1}}$(d).
To capture the exotic superconducting GLS at FEE, the contribution
of low-energy pairing induces a pseudospin energy shift given by $(1/2)\text{Tr}(\hat{\Delta}_{{\bf k}}^{++}(\hat{\Delta}_{{\bf k}}^{++})^{\dagger})=(81/32)k^{2}\Delta^{2}$
and $(1/2)\text{Tr}(\hat{\Delta}_{{\bf k}}^{--}(\hat{\Delta}_{{\bf k}}^{--})^{\dagger})=(9/32)k^{2}\Delta^{2}$.
Taking into account these shifts, the non-BdG spectrum for GLS above
and below $E=0$ are shown in Figs. $\text{\ref{Sup-Fig1}}$(e) and
(f), respectively. In this case, the exotic GLSs appear at $k_{1}=\tilde{k}=\sqrt{2\mu/(4\alpha+5\beta)}$.
Taking into account the pseudospin energy shifts, the excitation behaviors
around GLSs are consistent with full BdG spectrum in Fig. $\text{\ref{Sup-Fig1}}$(d).

\section{Constructing pairing in $O_h$ symmetry and \lowercase{$j$}$=3/2$ representation\label{App: Construction}}

We start by the density-density interaction decomposed in the pair
scattering formalism with the total intrinsic spin $S$ in the $j=3/2$
basis as
\begin{equation}
H_{\text{V}}=\sum_{{\bf k},{\bf k}^{\prime}}\sum_{S,m_{S}}V({\bf k}-{\bf k}^{\prime})\ b_{S,m_{S}}^{\dagger}\left({\bf k}\right)b_{S,m_{S}}\left({\bf k}^{\prime}\right),\label{TwoBody}
\end{equation}
where $b_{S,m_{S}}^{\dagger}\left({\bf k}\right)$ $\left[b_{S,m_{S}}\left({\bf k}^{\prime}\right)\right]$
creates (annihilates) a Cooper pair with intrinsic angular momentum
$S$ and spin magnetic quantum number $m_{S}$. The correspondence
between the Cooper pair operator and the two-electron state is given
by 
\begin{equation}
\!\!\!b_{S,m_{S}}^{\dagger}\!\left({\bf k}\right)\!\!=\!\!\!\!\!\sum_{m_{j_{1}},m_{j_{2}}}\!\!\!\!\langle j_{1},j_{2};m_{j_{1}},m_{j_{2}}|S,m_{S}\rangle c_{{\bf k},m_{j_{1}}}^{\dagger}\!c_{-{\bf k},m_{j_{2}}}^{\dagger}\!,\!\!\label{StoMj1Mj2}
\end{equation}
where $\langle j_{1},j_{2};m_{j_{1}},m_{j_{2}}|S,m_{S}\rangle$ is
the CGC connecting the two-electron state $|j_{1},j_{2};m_{j_{1}},m_{j_{2}}\rangle\equiv c_{{\bf k},m_{j_{1}}}^{\dagger}c_{-{\bf k},m_{j_{2}}}^{\dagger}|0\rangle$
with the Cooper pair state $|S,m_{S}\rangle=b_{S,m_{S}}^{\dagger}\left({\bf k}\right)|0\rangle$
in intrinsic total spin representation. Here, each electron has total
angular momentum $j$ with relative magnetic quantum number $m_{j}$.
For convenience, we can represent the Cooper pair operator in a compact
form with the aid of spin multipole matrices \citep{Theo1,Theo3,Mixing}
as
\begin{align}
b_{S,m_{S}}^{\dagger}\left({\bf k}\right) & =\hat{c}_{{\bf k}}^{\dagger}[\hat{\mathcal{S}}_{S,m_{S}}\hat{\mathcal{T}]}(\hat{c}_{-{\bf k}}^{\dagger})^{T},\nonumber \\
b_{S,m_{S}}\left({\bf k}^{\prime}\right) & =\left(\hat{c}_{-{\bf k}^{\prime}}\right)^{T}[\hat{\mathcal{S}}_{S,m_{S}}\hat{\mathcal{T}}]^{\dagger}\hat{c}_{{\bf k}^{\prime}},\label{CP operators}
\end{align}
where $\hat{\mathcal{T}}$ plays the role of Cooper pair symmetrization
and anti-symmetrization, and $\hat{\mathcal{S}}_{S,m_{S}}$ denotes
the rank-3 spherical spin multipole matrices corresponding to the
$j=$3/2 representation. Note that the $\hat{\mathcal{S}}_{S,m_{S}}$
have the properties of spin multipole moments since Cooper pairs are
formed with two charges. Hence, they can be classified as spin dipole,
quadruple, octupole moments, etc for $S=1,2,3,...$, respectively.
By comparing Eqs. (\ref{StoMj1Mj2}) and (\ref{CP operators}), we
can conclude that $\hat{\mathcal{S}}_{S,m_{S}}\hat{\mathcal{T}}$
is the matrix of CGCs relating the single particle Cooper pair state
to the two electron state. Therefore, to derive the multipole matrices,
we must find the highest weight matrix by setting $m_{S}=S$ and $\hat{\mathcal{S}}_{S,S}=\hat{\mathcal{C}}_{S,S}\hat{\mathcal{T}}^{-1}$
where $\mathcal{C}_{S,S}$ is the matrix composed of CGCs. Then, the
lower weight spin multipole matrices can be computed by the recursive
formula $[\hat{S}_{-},\hat{\mathcal{S}}_{S,m_{S}}]=\hbar\sqrt{S(S+1)-m_{S}(m_{S}-1)}\hat{\mathcal{S}}_{S,m_{S}-1}$
where $\hat{S}_{-}=\hat{S}_{x}-i\hat{S}_{y}$. Furthermore, the interaction
potential $V({\bf k}-{\bf k}^{\prime})$ can be expanded in terms
of spherical harmonics 
\begin{equation}
V({\bf k}-{\bf k}^{\prime})=\sum_{L,m_{L}}\frac{\alpha_{L}({\bf k},{\bf k}^{\prime})}{2L+1}Y_{L,m_{L}}(\mathsf{k})Y_{L,m_{L}}^{*}(\mathsf{k}^{\prime}).\label{Potential in sphearical}
\end{equation}
where the orbital axial angular momentum satisfies the condition $-L\leq m_{L}\leq L$
and the coefficient $\alpha_{L}({\bf k},{\bf k}^{\prime})$ can be
derived by $\alpha_{L}({\bf k},{\bf k}^{\prime})=\int d\varOmega\int d\varOmega^{\prime}\sum_{m_{L^{\prime}}}Y_{L^{\prime},m_{L^{\prime}}}^{*}(\mathsf{k})Y_{L^{\prime},m_{L^{\prime}}}(\mathsf{k}^{\prime})V({\bf k}-{\bf k}^{\prime})$.
Here, $\mathsf{k}$ denotes the norm of momentum vector. Note that
the orthogonality of spherical harmonics implies $\int d\varOmega Y_{L^{\prime},m_{L^{\prime}}}^{*}(\mathsf{k})Y_{L,m_{L}}(\mathsf{k})=\delta_{L,L^{\prime}}\delta_{m_{L},m_{L^{\prime}}}$
where $d\varOmega=\sin\left(\theta\right)d\theta d\varphi$. Inserting
Eqs. (\ref{CP operators}) and (\ref{Potential in sphearical}) into
Eq. (\ref{TwoBody}), this results in the interaction Hamiltonian
in the representation of $SO(3)$ symmetry 
\begin{align}
H_{V}^{(L,S)}\!\!=\!\!\!\!\sideset{}{^{\prime}}\sum & \Big[\hat{c}_{{\bf k}}^{\dagger}\big(Y_{L,m_{L}}(\mathsf{k})\hat{\mathcal{S}}_{S,m_{S}}\hat{\mathcal{T}}\big)(\hat{c}_{-{\bf k}}^{\dagger})^{T}\Big]\times\nonumber \\
 & \Big[(\hat{c}_{-{\bf k}^{\prime}})^{T}\big(Y_{L,m_{L}}(\mathsf{k}^{\prime})\hat{\mathcal{S}}_{S,m_{S}}\hat{\mathcal{T}}\big)^{\dagger}\hat{c}_{{\bf k}^{\prime}}\Big],\label{LS}
\end{align}
where $\sum^{\prime}\!=\sum_{{\bf k},{\bf k}^{\prime}}\sum_{S,m_{S}}\sum_{L,m_{L}}[\alpha_{L}({\bf k},{\bf k}^{\prime})/(2L+1)]$.
In our system, the total angular momentum $J$ is a good quantum number
due to presence of strong spin-orbit coupling. Thus, the density-density
interaction in Eq. (\ref{LS}) should be decomposed in the irreducible
representation of total angular momentum. Therefore, the function
matrices $Y_{L,m_{L}}(\mathsf{k})\hat{\mathcal{S}}_{S,m_{S}}$ can
be transformed into the total angular momentum $J$ basis by $Y_{L,m_{L}}(\mathsf{k})\hat{\mathcal{S}}_{S,m_{S}}=\sum_{J,m_{J}}\langle J,m_{J}|m_{L},m_{S}\rangle\hat{\mathcal{N}}_{J,m_{J}}^{S,L}(\mathsf{k})$
where $\langle J,m_{J}|m_{L},m_{S}\rangle$ denotes the CGC connecting
the Ket $|L,S;m_{L},m_{S}\rangle=|L,m_{L}\rangle\otimes|S,m_{S}\rangle$
with the Bra $\langle J,m_{J}|$. Note that the $\hat{\mathcal{N}}_{J,m_{J}}^{S,L}(\mathsf{k})$
indicates the total angular momentum multipole matrices. Eventually,
we arrive at the density-density interaction in the representation
of $J$ as

\begin{align}
H_{V}^{(J,S,L)}\!=\sideset{}{^{\prime\prime}}\sum & \Big[\hat{c}_{{\bf k}}^{\dagger}\big(\hat{\mathcal{N}}_{J,m_{J}}^{S,L}(\mathsf{k})\hat{\mathcal{T}}\big)(\hat{c}_{-{\bf k}}^{\dagger})^{T}\Big]\times\nonumber \\
 & \Big[(\hat{c}_{-{\bf k}^{\prime}})^{T}\big(\hat{\mathcal{N}}_{J,m_{J}}^{S,L}(\mathsf{k}^{\prime})\hat{\mathcal{T}}\big)^{\dagger}\hat{c}_{{\bf k}^{\prime}}\Big],\label{JLS}
\end{align}
where $\sum^{\prime\prime}=\sum_{{\bf k},{\bf k}^{\prime}}\sum_{J,m_{J}}\sum_{S,L}[\alpha_{L}({\bf k},{\bf k}^{\prime})/(2L+1)]$
and

\begin{equation}
\hat{\mathcal{N}}_{J,m_{J}}^{S,L}(\mathsf{k}) =\!\!\sum_{m_{L},m_{S}}\langle m_{L},m_{S}|J,m_{J}\rangle Y_{L,m_{L}}(\mathsf{k})\hat{\mathcal{S}}_{S,m_{S}}.
\end{equation}
This filters out magnetic orbital and axial spin angular momenta satisfying
$|m_{L}+m_{S}|=m_{J}$. Note that for $L=1$, we have $\alpha_{L=1}=4\pi|{\bf k}||{\bf k}^{\prime}|$.

In the presence of spherical symmetry, the gap functions are labeled
by an infinite number of IRs corresponding to $SO(3)$ symmetry, i.e.,
the total angular momentum $J$. However, crystals with cubic point
group structure have lower symmetry. Therefore, the corresponding
pairing instabilities and the corresponding Cooper pair operator must
be labeled by the IRs of $O_{h}$ symmetry. Thus, we need to derive
the cubic representation of $\hat{\mathcal{N}}_{J,m_{J}}^{S,L}(\mathsf{k})$.
This can be done by the following relation \citep{Bogo3}
\begin{equation}
\!\!\!\!\!\!\hat{N}_{\eta}^{J,S,L}\!(\mathsf{k})\!\!=\!\!\!\!\!\sideset{}{^{\prime\prime\prime}}\sum\!\!\!(\hat{\mathcal{O}}_{\eta}\hat{\mathcal{T}})_{\!m_{j_{1}}\!,m_{j_{2}}}\!\langle m_{j_{1}}\!,m_{j_{2}}|J,m_{J}\!\rangle\hat{\mathcal{N}}_{J\!,m_{J}}^{S,L}\!(\mathsf{k}),\!\!\!\!\label{CubicPair}
\end{equation}
where $\sum^{\prime\prime\prime}=\sum_{m_{j_{1}},m_{j_{2}}}\sum_{m_{J}}$
and $\eta$ is the basis label of cubic IRs and $\hat{\mathcal{O}}_{\eta}$
denotes the relative multipole basis matrices in cubic structure normalized
to identity, i.e., $\text{Tr}(\hat{\mathcal{O}}_{\eta}\hat{\mathcal{O}}_{\eta}^{\dagger})=1$.
The full information about $O_{h}$ pairing states and their relative
multipole matrices are listed in Table. \ref{Table1_OhIrreps}. Note
that Eq. (\ref{CubicPair}) shows full correspondence between cubic
point group symmetry and SO(3) symmetry. To obtain the density-density
interaction in the cubic field IR, we should replace $\hat{\mathcal{N}}_{J,m_{J}}^{S,L}(\mathsf{k})$
in Eq. (\ref{JLS}) with $\hat{N}_{\eta}^{J,S,L}(\mathsf{k})$ in
Eq. (\ref{CubicPair}). Performing a mean-field approximation with
the assumption that the electron pairs have zero center of momentum,
we obtain
\begin{equation}
\varUpsilon_{{\bf k}}^{\dagger}\varUpsilon_{{\bf k}^{\prime}}\thickapprox\!\langle\varUpsilon_{{\bf k}}^{\dagger}\rangle\varUpsilon_{{\bf k}^{\prime}}+\varUpsilon_{{\bf k}}^{\dagger}\langle\varUpsilon_{{\bf k}^{\prime}}\rangle+\langle\varUpsilon_{{\bf k}}^{\dagger}\rangle\langle\varUpsilon_{{\bf k}^{\prime}}\rangle,\label{Meanfield}
\end{equation}
where

\begin{align}
\varUpsilon_{{\bf k}}^{\dagger} & \equiv c_{{\bf k},m_{j_{1}}}^{\dagger}c_{-{\bf k},m_{j_{2}}}^{\dagger},\\
\varUpsilon_{{\bf k}^{\prime}} & \equiv c_{-{\bf k}^{\prime},m_{j_{3}}}c_{{\bf k}^{\prime},m_{j_{4}}}.
\end{align}
In the above relations, the independency of interaction on magnetic
quantum number requires $m_{j_{1}}=m_{j_{4}}$ and $m_{j_{2}}=m_{j_{3}}$.
This can be clearly seen by evaluating the matrix element of interaction
in two-electron state representation. The mean-field decomposition
in Eq. (\ref{Meanfield}) results in an effective single particle
formalism of a cubic pairing Hamiltonian in the channel $(\eta,J,S,L)$
\begin{equation}
H_{\eta}^{(J,S,L)}=\sum_{{\bf k}}\hat{c}_{{\bf k}}^{\dagger}\hat{\mathcal{H}}_{\eta}^{J,S,L}({\bf k})(\hat{c}_{-{\bf k}}^{\dagger})^{T}+h.c.,\label{H_cub_singleparticle}
\end{equation}
with 
\begin{equation}
\hat{\mathcal{H}}_{\eta}^{J,S,L}({\bf k})=|\boldsymbol{k}|^{L}\Delta_{\eta}^{J,S,L}\hat{N}_{\eta}^{J,S,L}(\mathsf{k})\hat{\mathcal{T}},\label{HcubicPairing}
\end{equation}
where $\Delta_{\eta}^{J,S,L}$ denotes the pairing strength defined
by 
\begin{equation}
\Delta_{\eta}^{J,S,L}=\sum_{{\bf k}}\langle G|\hat{c}_{-{\bf k}}^{T}(|\boldsymbol{k}|^{L}\ \hat{N}_{\eta}^{J,S,L}(\mathsf{k})\ \hat{\mathcal{T}})^{\dagger}\hat{c}_{{\bf k}}|G\rangle.\label{PairingAmplitude}
\end{equation}
with $|G\rangle$ being the superconducting ground state. Note that
in the Letter, we have taken the pairing strength as a (small) constant
$\Delta_{\eta}^{J,S,L}:=\Delta$ for all the stationary pairing states
in the weak-pairing limit.

\subsection{Symmetry properties of inter-band pairing}

In this section, we derive Eq. (9) of the main text. The Pauli exclusion
principle implies that a sign change in momentum space is accompanied
by exchanging the magnetic quantum numbers. This is encoded in $\hat{\mathcal{H}}_{\eta}^{J,S,L}(-{\bf k})=-(\hat{\mathcal{H}}_{\eta}^{J,S,L}({\bf k}))^{T}$,
therefore, $\hat{N}_{\eta}^{J,S,L}(-\mathsf{k})\hat{\mathcal{T}}=-(\hat{N}_{\eta}^{J,S,L}(\mathsf{k})\hat{\mathcal{T}})^{T}$.
Using these relations, we can directly find the symmetry of the inter-band
pairing Hamiltonian by projecting Eq. (\ref{HcubicPairing}) into
the inter-band subspace as 
\begin{align}
\hat{\Delta}_{-{\bf k}}^{+-} & =\hat{V}_{-{\bf k}}^{+\dagger}\left(|\boldsymbol{k}|^{L}\Delta_{\eta}^{J,S,L}\hat{N}_{\eta}^{J,S,L}(-\mathsf{k})\hat{\mathcal{T}}\right)\hat{V}_{{\bf k}}^{-\dagger^{T}}\nonumber \\
 & =-\hat{V}_{-{\bf k}}^{+\dagger}\left(|\boldsymbol{k}|^{L}\Delta_{\eta}^{J,S,L}[\hat{N}_{\eta}^{J,S,L}(\mathsf{k})\hat{\mathcal{T}}]^{T}\right)\hat{V}_{{\bf k}}^{-\dagger^{T}}\nonumber \\
 & =-[\hat{\Delta}_{{\bf k}}^{-+}]^{T},
\end{align}
where $\hat{V}_{-{\bf k}}^{\pm}=\hat{V}_{{\bf k}}^{\pm}$ due to inversion
symmetry. In the next sections, we present the explicit form of all
allowed symmetry stationary $s$- and $p$-wave cubic pairings.
\begin{center}
\begin{table}
\begin{tabular}{ccccc}
\hline 
$J$ & $O_{h}$ & $\eta$ & $\mathcal{O}_{\eta}\left(J\right)$ & $\text{Stationary state}$\tabularnewline
\hline 
$0$ & $A_{1g(u)}$ & $I\ (f(\boldsymbol{r}))$ & $I_{4}$ & $\checkmark$\tabularnewline
$1$ & $T_{1u}$ & $x$ & $J_{x}$ & $\times$\tabularnewline
 &  & $y$ & $J_{y}$ & $\times$\tabularnewline
 &  & $z$ & $J_{z}$ & $\checkmark$\tabularnewline
$2$ & $E_{g,u}$ & $3z^{2}-r^{2}$ & $3J_{z}^{2}-\boldsymbol{J}^{2}$ & $\checkmark$\tabularnewline
 &  & $x^{2}-y^{2}$ & $J_{x}^{2}-J_{y}^{2}$ & $\checkmark$\tabularnewline
 & $T_{2g,u}$ & $xy$ & $\lceil J_{x}J_{y}\rfloor$ & $\times$\tabularnewline
 &  & $zx$ & $\lceil J_{z}J_{x}\rfloor$ & $\times$\tabularnewline
 &  & $yz$ & $\lceil J_{y}J_{z}\rfloor$ & $\times$\tabularnewline
$3$ & $A_{2u}$ & $xyz$ & $\lceil J_{x}J_{y}J_{z}\rfloor$ & $\checkmark$\tabularnewline
 &  & $x^{3}$ & $J_{x}^{3}$ & $\times$\tabularnewline
 & $T_{1u}$ & $y^{3}$ & $J_{y}^{3}$ & $\times$\tabularnewline
 &  & $z^{3}$ & $J_{z}^{3}$ & $\times$\tabularnewline
 & $T_{2u}$ & $z(x^{2}-y^{2})$ & \ \ $\lceil J_{z}(J_{x}^{2}-J_{y}^{2})\rfloor$ & $\times$\tabularnewline
 &  & $x(y^{2}-z^{2})$ & \ \ $\lceil J_{x}(J_{y}^{2}-J_{z}^{2})\rfloor$ & $\times$\tabularnewline
 &  & $y(z^{2}-x^{2})$ & \ \ $\lceil J_{y}(J_{z}^{2}-J_{x}^{2})\rfloor$ & $\times$\tabularnewline
\hline 
\end{tabular}\caption{Time-reversal pairing states in cubic point group symmetry. The first
and second columns identify the correspondence between total angular
momentum of pairing states and IRs of the $O_{h}$ symmetry \citep{GroTinkham,GroDress}.
The real space basis of each IR is denoted in the third column with
their corresponding $J$ basis in the forth column. Here, the square
brackets $\lceil...\rfloor$ symmetrize the multipole basis operator
$\lceil\hat{A}\hat{B}\rfloor=(\hat{A}\hat{B}+\hat{B}\hat{A})/2!$
and $\lceil\hat{A}\hat{B}\hat{C}\rfloor=(\hat{A}\hat{B}\hat{C}+\hat{A}\hat{C}\hat{B}+\hat{B}\hat{C}\hat{A}+\hat{B}\hat{A}\hat{C}+\hat{C}\hat{A}\hat{B}+\hat{C}\hat{B}\hat{A})/3!$.
In the last column, $\checkmark$ ($\times$) implies that the pairing
state is (is not) the stationary state of the free energy \citep{Theo3}.}
\label{Table1_OhIrreps}
\end{table}
\par\end{center}

\subsection{s-wave pairing in $O_{h}$ symmetry\label{App: B}}

The pairing state with s-wave orbital angular momentum $L=0$ is allowed
in the even-parity $A_{1g}$ and $E_{g}$ channels. Here, we derive
the cubic pairing states corresponding to the relative allowed symmetry
quantum numbers $(J,S,L)$. The even-parity $A_{1g}$ state in the
channel $(0,0,0)$ and $E_{g}$ states in the channel $(2,2,0)$ are
given by 
\begin{align}
\hat{N}_{I}^{0,0,0}(\mathsf{k}) & =\hat{\mathcal{N}}_{0,0}^{0,0}(\mathsf{k}),\\
\hat{N}_{3z^{2}-r^{2}}^{2,2,0}(\mathsf{k}) & =\hat{\mathcal{N}}_{2,0}^{2,0}(\mathsf{k}),\label{App:Eg-3z2:220}\\
\hat{N}_{x^{2}-y^{2}}^{2,2,0}(\mathsf{k}) & =\frac{1}{\sqrt{2}}\left(\hat{\mathcal{N}}_{2,-2}^{2,0}(\mathsf{k})+\hat{\mathcal{N}}_{2,2}^{2,0}(\mathsf{k})\right).
\end{align}
Inserting the above relations into Eq. (\ref{HcubicPairing}), this
results in the explicit matrix formalism of the pairing Hamiltonians
as 
\begin{align}
\hat{\mathcal{H}}_{I}^{0,0,0}({\bf k}) & =\Delta_{I}^{0,0,0}\left(\begin{array}{cccc}
0 & 0 & 0 & 1\\
0 & 0 & -1 & 0\\
0 & 1 & 0 & 0\\
-1 & 0 & 0 & 0
\end{array}\right),\label{000}
\end{align}
\begin{align}
\hat{\mathcal{H}}_{3z^{2}-r^{2}}^{2,2,0}({\bf k}) & =\Delta_{3z^{2}-r^{2}}^{2,2,0}\left(\begin{array}{cccc}
0 & 0 & 0 & 1\\
0 & 0 & 1 & 0\\
0 & -1 & 0 & 0\\
-1 & 0 & 0 & 0
\end{array}\right),\label{Eg220-3z2mr2}
\end{align}
\begin{align}
\hat{\mathcal{H}}_{x^{2}-y^{2}}^{2,2,0}({\bf k}) & =\Delta_{x^{2}-y^{2}}^{2,2,0}\left(\begin{array}{cccc}
0 & 1 & 0 & 0\\
-1 & 0 & 0 & 0\\
0 & 0 & 0 & 1\\
0 & 0 & -1 & 0
\end{array}\right).\label{Eg220-x2my2}
\end{align}

\subsection{$p$-wave pairing in $O_{h}$ symmetry\label{App: P-wave}}

The $p$-wave gap functions implies that $L=1$. Therefore, the superconducting
gap functions depend linearly on momentum. Since the orbital angular
momentum is odd, consequently, the intrinsic spin part of Cooper pairs
should be odd due to Fermi statistics. Therefore, the p-wave gap functions
are odd in momentum implying that $\hat{\mathcal{H}}_{\eta}^{J,S,L}(-{\bf k})=-\hat{\mathcal{H}}_{\eta}^{J,S,L}({\bf k})$.
Taking into account the condition $|S-L|\leq J\leq|S+L|$, the odd
parity p-wave Cooper pairing can possess either spin dipole structure
$S=1$ or spin octupole structure $S=3$. Hence, Cooper pairs can
have singlet $J=0$, triplet $J=1$, and quintet $J=2$ total angular
momenta for spin dipole moment. Also, quintet $J=2$ and septet $J=3$
total angular momenta correspond to pairing with spin octupole structure.
It is worth mentioning that the triplet and septet pairings can only
happen in the $p$-wave channel. In the following, we obtain the explicit
matrix formalism of Hamiltonians describing $p$-wave pairings in
$O_{h}$ point group symmetry. This can be done by inserting Eq. (\ref{CubicPair})
into Eq. (\ref{HcubicPairing}).

\subsubsection{Singlet state $J=0$}

Here, we derive the odd-parity pairing state $A_{1u}$. The symmetry
constraint allows for the channel $(0,1,1)$. The $J$ representation
of the $A_{1u}$ state and the full matrix formalism of pairing results
in
\begin{align}
 & \hat{N}_{f(\boldsymbol{r})}^{0,1,1}(\mathsf{k})=\hat{\mathcal{N}}_{0,0}^{1,1}(\mathsf{k}),\nonumber \\
 & \!\!\hat{\mathcal{H}}_{f(\boldsymbol{r})}^{0,1,1}({\bf k})\!\!=\!\!\Delta_{f(\boldsymbol{r})}^{0,1,1}\!\!\left(\!\!\begin{array}{cccc}
0 & 0 & -\frac{\sqrt{3}}{2}k_{-} & \frac{3k_{z}}{2}\\
0 & k_{-} & -\frac{1}{2}k_{z} & \frac{\sqrt{3}}{2}k_{+}\\
-\frac{\sqrt{3}}{2}k_{-} & -\frac{1}{2}k_{z} & -k_{+} & 0\\
\frac{3k_{z}}{2} & \frac{\sqrt{3}}{2}k_{+} & 0 & 0
\end{array}\!\!\!\!\right)\!\!.\!\!\!\!\label{App:A1u-011}
\end{align}

\subsubsection{Triplet state $J=1$}

In the cubic field, the $J=1$ state is labeled by the $T_{1u}$ IR
which is a three-fold degenerate state, each denoted by the basis
$\eta={x,y,z}$. Note that only $\eta=z$ is a stationary state of
the free energy preserving time-reversal symmetry \citep{InertStates,Theo3}.
Hence, we focus on it . The total angular momentum representation
of this state which lies in the channel $(1,1,1)$ is 
\begin{equation}
\hat{N}_{z}^{1,1,1}(\mathsf{k})=\hat{\mathcal{N}}_{1,0}^{1,1}(\mathsf{k}).\label{spinTripletStates}
\end{equation}
The full matrix of the odd-parity triplet pairing takes the form 
\begin{equation}
\hat{\mathcal{H}}_{z}^{1,1,1}({\bf k})\!=\!\Delta_{z}^{1,1,1}\!\!\left(\begin{array}{cccc}
0 & 0 & k_{-} & 0\\
0 & -\frac{2}{\sqrt{3}}k_{-} & 0 & k_{+}\\
k_{-} & 0 & -\frac{2}{\sqrt{3}}k_{+} & 0\\
0 & k_{+} & 0 & 0
\end{array}\!\!\right).\!\!\!\label{=00003D1, z}
\end{equation}

\subsubsection{Quintet state $J=2$}

Here, we derive the odd-parity pairing states with $E_{u}$ and $T_{2u}$
IR corresponding to states with quintet total angular momentum. It
is worth mentioning that the $J$ representation of these IR are the
same as Eqs. (\ref{Eg220-3z2mr2})-(\ref{Eg220-x2my2}). We represent
them for $(S=1,L=1)$ and $(S=3,L=1)$ channels. The pairing Hamiltonian
of the former channel takes the form 
\begin{align}
 & \hat{\mathcal{H}}_{3z^{2}-r^{2}}^{2,1,1}({\bf k})\!\!=\!\!\Delta_{3z^{2}-r^{2}}^{2,1,1}\!\!\left(\!\!\!\begin{array}{c@{\hskip-1pt}c@{\hskip-1pt}c@{\hskip-1pt}ccccccccccccccccccccccccccccccccccccccccccc}
0 & 0 & \frac{\sqrt{3}}{2}k_{-} & 3k_{z}\\
0 & -k_{-} & -k_{z} & -\frac{\sqrt{3}}{2}k_{+}\\
\frac{\sqrt{3}}{2}k_{-} & -k_{z} & k_{+} & 0\\
3k_{z} & -\frac{\sqrt{3}}{2}k_{+} & 0 & 0
\end{array}\!\!\right)\!\!,\!\!\!\!\label{Eu:211,3z^2-r^2}\\
 & \hat{\mathcal{H}}_{x^{2}-y^{2}}^{2,1,1}({\bf k})\!\!=\!\!\Delta_{x^{2}-y^{2}}^{2,1,1}\!\!\left(\!\!\begin{array}{cccc}
0 & 0 & -k_{+} & 0\\
0 & \frac{2}{\sqrt{3}}k_{+} & 0 & k_{-}\\
-k_{+} & 0 & -\frac{2}{\sqrt{3}}k_{-} & 0\\
0 & k_{-} & 0 & 0
\end{array}\!\!\right)\!\!.\!\!\label{Eu:211,x^2-y^2}
\end{align}
Moreover, the quintet Hamiltonians with spin octupole $S=3$ structures
are given by 
\begin{align}
 & \!\!\hat{\mathcal{H}}_{3z^{2}-r^{2}}^{2,3,1}({\bf k})\!\!=\!\!\Delta_{3z^{2}-r^{2}}^{2,3,1}\!\!\left(\!\!\begin{array}{cccc}
0 & 0 & \frac{1}{\sqrt{3}}k_{-} & \frac{-1}{2}k_{z}\\
0 & k_{-} & \frac{-3}{2}k_{z} & \frac{-1}{\sqrt{3}}k_{+}\\
\frac{1}{\sqrt{3}}k_{-} & \frac{-3}{2}k_{z} & -k_{+} & 0\\
\frac{-1}{2}k_{z} & \frac{-1}{\sqrt{3}}k_{+} & 0 & 0
\end{array}\!\!\!\!\!\right)\!\!,\!\!\!\!\label{App:3z2-r2: 231}
\end{align}
\begin{align}
 & \!\!\!\!\hat{\mathcal{H}}_{x^{2}-y^{2}}^{2,3,1}\!({\bf k})\!\!=\!\!\Delta_{x^{2}-y^{2}}^{2,3,1}\!\!\left(\!\!\!\begin{array}{c@{\hskip-1pt}c@{\hskip-1pt}cccccccccccccccccccccccccccccccccccccccccccccccccc}
5\sqrt{3}k_{-} & -5k_{z} & -k_{+} & 0\\
-5k_{z} & -\!\sqrt{3}k_{+} & 0 & k_{-}\\
-k_{+} & 0 & \sqrt{3}k_{-} & -5k_{z}\\
0 & k_{-} & -5k_{z} & -5\sqrt{3}k_{+}
\end{array}\!\!\!\right)\!\!.\!\!\!\!\!\!\label{App:x^2-y^2: 231}
\end{align}

\subsubsection{Septet state $J=3$\label{App: J3 Cubic Pairing}}

The $p$-wave septet $J=3$ state in cubic representation decomposes
into $A_{2u}+T_{1u}+T_{2u}$ IR \citep{GroTinkham,GroDress}. In this
case, the Cooper pairs have only intrinsic spin octupole structure
$S=3$. The $A_{2u}$ state \citep{Brydon} is a stationary state
\citep{Theo3} and its matrix Hamiltonian is given by 
\begin{equation}
\hat{N}_{xyz}^{3,3,1}(\mathsf{k})=\frac{1}{i\sqrt{2}}\left(\hat{\mathcal{N}}_{3,2}^{3,1}(\mathsf{k})-\hat{\mathcal{N}}_{3,-2}^{3,1}(\mathsf{k})\right),
\end{equation}
\begin{align}
\!\!\hat{\mathcal{H}}_{xyz}^{3,3,1}({\bf k})\!\!=\!\!\Delta_{xyz}^{3,3,1}\!\!\left(\!\!\!\begin{array}{cccc}
\frac{3}{4}k_{-} & \frac{\sqrt{3}}{2}k_{z} & \frac{\sqrt{3}}{4}k_{+} & 0\\
\frac{\sqrt{3}}{2}k_{z} & \frac{3}{4}k_{+} & 0 & -\frac{\sqrt{3}}{4}k_{-}\\
\frac{\sqrt{3}}{4}k_{+} & 0 & -\frac{3}{4}k_{-} & \frac{\sqrt{3}}{2}k_{z}\\
0 & -\frac{\sqrt{3}}{4}k_{-} & \frac{\sqrt{3}}{2}k_{z} & -\frac{3}{4}k_{+}
\end{array}\!\!\!\right)\!\!.\!\!\!\!\label{xyz}
\end{align}

\section{Multipole matrices in \lowercase{$j$}$=3/2$ representation \label{App: Spin multipole}}

The multipole matrix for $S=1$ and $S=3$ with the highest $m_{S}$
quantum numbers are 
\begin{align}
\!\!\hat{\mathcal{S}}_{1,1}\!\!=\!\!\left(\!\!\begin{array}{cccc}
0 & \frac{-\sqrt{3}}{\sqrt{10}} & 0 & 0\\
0 & 0 & \frac{-\sqrt{2}}{\sqrt{5}} & 0\\
0 & 0 & 0 & \frac{-\sqrt{3}}{\sqrt{10}}\\
0 & 0 & 0 & 0
\end{array}\!\!\right)\!,\ \hat{\mathcal{S}}_{3,3}\!\!=\!\!\left(\!\!\begin{array}{cccc}
0 & 0 & 0 & -1\\
0 & 0 & 0 & 0\\
0 & 0 & 0 & 0\\
0 & 0 & 0 & 0
\end{array}\!\!\right)\!\!.\!\label{S1S3}
\end{align}
Furthermore, the spherical harmonic for $L=1$ are $|{\bf k}|Y_{1,\pm1}=\mp\sqrt{3/8\pi}k_{\pm}$
and $|{\bf k}|Y_{1,0}=\sqrt{3/4\pi}k_{z}$. The $\hat{J}_{i}$ ($\hat{S}_{i}$)
matrices with $i\in\{x,y,z\}$ in $j=3/2$ basis are given\textcolor{red}{{}
}by
\begin{align}
\hat{J}_{x} & =\hat{S}_{x}=\frac{\hbar}{2}\left(\begin{array}{cccc}
0 & \sqrt{3} & 0 & 0\\
\sqrt{3} & 0 & 2 & 0\\
0 & 2 & 0 & \sqrt{3}\\
0 & 0 & \sqrt{3} & 0
\end{array}\right),\\
\hat{J}_{y} & =\hat{S}_{y}=\frac{\hbar}{2}\left(\begin{array}{cccc}
0 & -i\sqrt{3} & 0 & 0\\
i\sqrt{3} & 0 & -i2 & 0\\
0 & i2 & 0 & -i\sqrt{3}\\
0 & 0 & i\sqrt{3} & 0
\end{array}\right),\\
\hat{J}_{z} & =\hat{S}_{z}=\frac{\hbar}{2}\left(\begin{array}{cccc}
3 & 0 & 0 & 0\\
0 & 1 & 0 & 0\\
0 & 0 & -1 & 0\\
0 & 0 & 0 & -3
\end{array}\right).
\end{align}
\clearpage

\begin{thebibliography}{75}
\bibitem [{cm7()}]{cm7}%
  \BibitemOpen
  \href@noop {} {\ }\bibinfo {note} {\!\!The analogous model for $(-+)$
  subspace can be derived by substituting $(+)\leftrightarrow(-)$ in Eq.
  \eqref{Heff+-}.}\BibitemShut {Stop}%
\bibitem [{cm9()}]{cm9}%
  \BibitemOpen
  \href@noop {} {\ }\bibinfo {note} {\!\!\!We expand the electron-electron
  interaction through the orthogonal basis set of cubic point group symmetry.
  In this regard, the expanded two-body interaction includes many terms each
  distinguished by the basis label of cubic symmetry and Cooper pair quantum
  numbers. We investigate each term individually.}\BibitemShut {Stop}%
\bibitem [{cm4()}]{cm4}%
  \BibitemOpen
  \href@noop {} {\ }\bibinfo {note} {\!\!For cubic anisotropy in the normal-state $\gamma\neq\beta$, our results are still valid.}\BibitemShut {Stop}%
\bibitem [{ChanNaming()}]{ChanNaming}%
  \BibitemOpen
  \href@noop {} {\ }\bibinfo {note} {\!\!The pairing channel nomenclature is based on $J$ quantum number of Cooper pairs due to presence of strong spin-orbit coupling.}\BibitemShut {Stop}%
\bibitem [{Eqi()}]{EqiDirec}%
  \BibitemOpen
  \href@noop {} {\ }\bibinfo {note} {\!\!\!There are three (eight) equivalent
  directions for the proper two(three)-fold rotation $C_{2(3)}$ in cubic point
  group symmetry. The equivalent direcions for the $C_{2}$ rotation are
  $[1,0,0]$, $[0,1,0]$ and $[0,0,1]$. For the $C_{3}$ rotation, the equivalent
  directions are $[1,1,1]$, $[\bar{1},1,1]$ , $[1,\bar{1},1]$, $[1,1,\bar{1}]$,
  $[\bar{1},\bar{1},1]$, $[1,\bar{1},\bar{1}]$, $[\bar{1},1,\bar{1}]$ and
  $[\bar{1},\bar{1},\bar{1}]$, where $\bar{1}\equiv-1$.}\BibitemShut {Stop}%
\bibitem [{cm()}]{cm}%
  \BibitemOpen
  \href@noop {} {\ }\bibinfo {note} {\!\!In Fig. \ref{Fig2}(a), the two-fold
  degeneracy along the $[0,0,1]^*$ direction is
  not lifted in the presence of inversion breaking ASOC term. This degeneracy
  is protected by mirror reflection symmetry and can be witnessed in
  time-reversal symmetric $T_d$ crystals \cite{InvBreak1}.}\BibitemShut {Stop}%
\end{thebibliography}
\end{document}